\documentclass[fleqn,usenatbib]{mnras}

\usepackage[T1]{fontenc}

\DeclareRobustCommand{\VAN}[3]{#2}
\let\VANthebibliography\thebibliography
\def\thebibliography{\DeclareRobustCommand{\VAN}[3]{##3}\VANthebibliography}

\usepackage{graphicx}	
\usepackage{amsmath,amssymb,amstext}	
\usepackage{multirow}
\usepackage{upgreek}
\usepackage{siunitx}

\usepackage{hyperref}
\usepackage[normalem]{ulem}
\hypersetup{colorlinks = true, citecolor = {blue}, linkcolor= {red}}

\def\PGPU{$\varphi-$GPU}

\def\gapprox{\;\rlap{\lower 3.0pt                       
        \hbox{$\sim$}}\raise 2.5pt\hbox{$>$}\;}
\def\lapprox{\;\rlap{\lower 3.1pt                       
        \hbox{$\sim$}}\raise 2.7pt\hbox{$<$}\;}

\newcommand{\be}{ \begin{equation} }
\newcommand{\ee}{\end{equation}}

\newcommand{\ben}{\begin{enumerate}}
\newcommand{\een}{\end{enumerate}}

\newcommand{\orcid}[1]{\href{https://orcid.org/#1}{\protect\includegraphics[width=8pt]{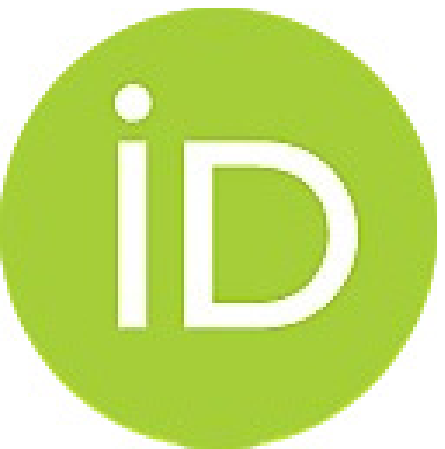}}}

\usepackage{newtxtext,newtxmath}

\title[NGC 6240 binary evolution based on Chandra data]{
NGC~6240 Supermassive Black Hole Binary dynamical evolution based on \textit{Chandra} data}

\author[M. Sobolenko et al.]{
M.~Sobolenko\orcid{0000-0003-0553-7301},$^{1}$\thanks{Contact e-mail:
{\href{mailto:sobolenko@mao.kiev.ua}{sobolenko@mao.kiev.ua}}}
O.~Kompaniiets\orcid{0000-0002-8184-6520}$^{1,2}$, 
P.~Berczik\orcid{0000-0003-4176-152X}$^{3,4,1}$, 
V.~Marchenko\orcid{0000-0002-7175-1923}$^{5}$, 
A.~Vasylenko$^{1}$, 
\newauthor
\;E.~Fedorova\orcid{0000-0002-8882-7496}$^{6,7,8}$,
B.~Shukirgaliyev\orcid{0000-0002-4601-7065}$^{9,10,11}$
\\
$^{1}$Main Astronomical Observatory, National Academy of Sciences of Ukraine, 27 Akademika Zabolotnoho St, 03143 Kyiv, Ukraine\\
$^{2}$Institute of Physics, National Academy of Sciences of Ukraine, 46 av. Nauky, 03028 Kyiv, Ukraine \\
$^{3}$Astronomisches Rechen-Institut, Zentrum f\"ur Astronomie, University of Heidelberg, M\"onchhofstrasse 12-14, 69120 Heidelberg, Germany\\
$^{4}$Konkoly Observatory, Research Centre for Astronomy and Earth Sciences, E\"otv\"os Lor\'and Research Network (ELKH),\\ 
\;\;MTA Centre of Excellence, Konkoly Thege Mikl\'os \'ut 15-17, 1121 Budapest, Hungary\\
$^{5}$Astronomical Observatory, Jagiellonian University, 171 ul. Orla, 30-244 Krakow, Poland\\
$^{6}$Astronomical Observatory, National Taras Shevchenko University of Kyiv, 3 Observatorna str., 04053 Kyiv, Ukraine \\
$^{7}$INAF-Osservatorio Astronomico di Roma, via Frascati 33, I-00078 Monte Porzio Catone, Italy \\
$^{8}$INAF-Osseravatorio Astrofisico di Catania, Universita di Catania, 95123 Catania, Italy\\
$^{9}$Energetic Cosmos Laboratory, Nazarbayev University, 53 Kabanbay Batyr ave, 010000 Nur-Sultan, Kazakhstan \\
$^{10}$Fesenkov Astrophysical Institute, 23 Observatory str, 050020 Almaty, Kazakhstan \\
$^{11}$Al-Farabi Kazakh National University, 71 Al-Farabi ave, 050020 Almaty, Kazakhstan \\
}

\date{Accepted XXX. Received YYY; in original form ZZZ}

\pubyear{2022}

\begin{document}
\label{firstpage}
\pagerange{\pageref{firstpage}--\pageref{lastpage}}
\maketitle

\begin{abstract}
The main idea of our research is to estimate the physical coalescence time of the double supermassive black hole (SMBH) system in the centre of NGC~6240 based on the X-ray observations from the \textit{Chandra} space observatory. The spectra of the Northern and Southern nuclei were fitted by spectral models from Sherpa and both presented the narrow component of the Fe~K$\alpha$ emission line. It enabled us to apply the spectral model to these lines and to find relative offset $\approx0.02$~keV. The enclosed dynamical mass of the central region of NGC~6240 with radius 1~kpc was estimated $\approx 2.04\times 10^{11} \rm\; M_{\odot}$. These data allowed us to carry on the high resolution direct N-body simulations with Newtonian and post-Newtonian (up to $2.5\mathcal{PN}$ correction) dynamics for this particular double SMBH system. As a result, from our numerical models we approximated the central SMBH binary merging time for the different binary eccentricities. In our numerical parameters range the upper limit for the merging time, even for the very small eccentricities, is still below $\approx70$~Myr. Gravitational waveforms and amplitude-frequency pictures from such events can be detected using Pulsar Timing Array (PTA) projects at the last merging phase. 
\end{abstract}

\begin{keywords}
galaxies:~active -- galaxies:~kinematics~and~dynamics -- galaxies:~individual:~NGC6240 --
X-rays:~galaxies -- black~hole~physics -- gravitational~waves
\end{keywords}



\section{Introduction} \label{sec:Intro}
The model of hierarchical galaxy evolution predicts galactic mergers \citep{White1978,Blumenthal1984,Kauffmann1999,Menci2002,Dobrycheva2018,Zoldan2019}. Since the most observed galactic nuclei harbour the supermassive black holes (SMBHs) in their centre \citep{Richstone1998,Ferrarese2000,Barausse2012,Vavilova2015}, the mergers of galaxies nearly always lead to the formation of the binary system of corresponding central SMBHs \citep{Kormendy1995}. Their evolution in the interacting galaxies can be described by three basic stages \citep{Begelman1980}.

In gas-free (dry merging) system the SMBHs become gravitational bound and create SMBH binary (SMBHB) when the semimajor axis approximately equals SMBHB influence radius. It is a sphere radius that contains within the stellar mass equal to double black hole (BH) mass. The duration of this stage depends on the efficiency of the dynamical friction, but the system definitely forms a pc-scale SMBHB. Afterwards, the SMBHB separation shrinks due to the combined effect of dynamical friction and gravitational slingshot. When the latter process becomes dominating the binary reaches the hardening phase with a semimajor axis \citep{Quinlan1996,Yu2002}:
\begin{equation}
a_{\rm h} \equiv \frac{G\mu}{4\sigma^{2}_{\ast}},
\end{equation}
where $G$ is the gravitational constant, the  binary reduced mass is $\mu=M_{\rm BH1}M_{\rm BH2}/(M_{\rm BH1}+M_{\rm BH2})$ with primary and secondary BHs' masses $M_{\rm BH1}$ and $M_{\rm BH2}$ respectively, and $\sigma_{\ast}$ is the velocity dispersion. The last merging stage is starting as the rapid coalescence of SMBHB via emission of gravitation waves (GWs) \citep{Peters1963,Peters1964a,Peters1964b,Haehnelt1994,Milosavljevic2003,Wyithe2003}. After coalescence, a single formed SMBH is kicked from the merger remnant centre and is observed as recoiling SMBH \citep{Campanelli2007,Choi2007,Gonzalez2007}. The accompanying emission of GWs is equivalently taking away up to the 10 per cent of total rest-mass of binary system \citep{Reisswig2009}.

SMBHB evolution can be stalled between hardening and GW phases due to depletion of loss cone and merging time is becoming above Hubble time. The so-called `final parsec' problem \citep{Milosavljevic2003b} occurs for idealised systems and can be solved in numerical simulations using the self-consistent equilibrium axisymmetric galaxy model \citep{Berczik2006,Preto2011}, using particles that have multiple encounters with central BHs \citep{Avramov2021} or using massive perturbers in loss cone \citep{Perets2007}. Also, the presence of gas in interacting systems (wet merging) plays a significant but unpredictable role, that can decrease or increase the SMBHB merging time depending on system parameters \citep[][for recent studies of gas and stars coinfluence see \citealt{Bortolas2021}]{Cuadra2009,Lodato2009,Maureira2018}.

The natural way to search for such SMBHs is by looking at dual or binary active galactic nuclei (AGNs) \citep{Husemann2020}. Except for SMBH, the AGN also contains major components such as the accretion disk around the BH and molecular torus \citep[e.g.][]{Ricci2014,Vasylenko2018,Grobner2020, Kompaniiets2020}. Accretion onto  SMBH  is accompanied by converting the gravitational potential energy to the observed radiation, spanning the entire electromagnetic spectrum. Most of this energy dissipates in the innermost few gravitational radii leading to the bright  X-ray emission.

X-ray radiation of AGN commonly is explained by thermal Comptonization of the soft ultraviolet (UV) radiation, produced by the inner parts of the accretion disk in a medium of `hot' electrons around SMBH known as the corona \citep{Haardt1991, Haardt1993}. This radiation (called the primary emission) typically is described by a power-law model and an exponential cut-off at high energies where emission quickly roll-overs \citep{Rybicki1979}. Additionally to the continuum is specified the important reflected component, which is the reprocessed primary emission by a cold neutral circumnuclear medium (molecular torus or outer regions of the accretion disk). It is observed as a `reflection hump' at $\sim$20--30~keV and emission in  Fe~K$\alpha$ line at around 6.4~keV \citep[e.g.][]{Matt1991,Mushotzky1993}. Due to a  combination of abundances and fluorescent yield, the neutral Fe~K$\alpha$ at 6.4~keV is typically the strongest emission line seen in AGN's X-ray spectra. If we found the energy difference for the observed Fe~K$\alpha$ lines we can assume that this shift is due to relative motion between two nuclei at the late stage. That gives the possibility to estimate the mass, enclosed within the common orbit of the binary system (i.e., dynamical mass).

One of the most prominent dual AGN candidates is nearby ultraluminous infrared galaxy (ULIRG) NGC~6240 ($z=0.0243$, $D_{\rm L}=111.2$~Mpc\footnote{\url{https://ned.ipac.caltech.edu/}}) that contains two heavily obscured Compton-thick nuclei separated by $\sim$1\farcs8 \citep{Gerssen2004}. Multiple multiwavelength observations unfold complex morphological structure and confirm that it is in an active merging state \citep{Pasquali2003}. Clearly visible by {\it Hubble Space Telescope} ({\it HST}) irregular elongated morphology of this galaxy is often referred as `butterfly' or `lobster-shaped' \citep{Muller2018}. NGC~6240 is observed as the AGN in X-ray \citep{Komossa2003,Puccetti2016,Fabbiano2020}. It shows intensive starburst activity \citep{Barger1998}, supernova explosions of young hot stars \citep{Maza2010} and contains H$_{2}$O maser \citep{Hagiwara2002,Hagiwara2003}. Another interesting property of this galaxy is the presence of the significant amount of dust surrounding the nucleus that causes its high infrared (IR) luminosity 
\citep[$\sim$10$^{12}\rm\;{L_\odot}$; see][]{Sanders2003,Iono2007}\footnote{$L_{\rm IR}$ is the 8--100~$\upmu$m luminosity}.
The Multi-Element Radio Linked Interferometer Network (MERLIN) observations at 1.4~Ghz and 5~Ghz revealed two compact radio sources in the nuclei of NGC~6240 \citep{Beswick2001}. Followed-up high-resolution observations using Very Long Baseline Array (VLBA) and Very Long Baseline Interferometry (VLBI) detected a more complex structure of the central region with several radio sources. Two of the radio sources, namely N1 (Northern nucleus, further N) and S (or N2, Southern nucleus), matched with compact X-ray sources. The N nucleus may be clearly classified as AGN according to the characteristics in the radio band. The S nucleus spectrum contains composite emission from the AGN and circumnuclear starburst/supernova remnants \citep{Gallimore2004,Hagiwara2011}. Recently, the results by \cite{Kollatschny2020} and \cite{Fabbiano2020} about the double structure of the S-nucleus are under discussion.

The SMBH mass of the S nucleus lies in the range $(0.87-2.0)\times10^9\rm\;M_{\odot}$ obtained from the high resolution stellar kinematic results \citep{Medling2011}. Using $K$-band data from Very Large Telescope (VLT) and classical $M_{\rm BH}$--$\sigma$ relation \citep{Tremaine2002}, the N and S nucleus SMBH masses were estimated as $(1.4\pm0.4)\times10^8\rm\;M_{\odot}$ and  $(2.0\pm0.4)\times10^{8}\rm\;M_{\odot}$, respectively \citep{Engel2010}. Later, \cite{Treister2020} noticed that used by \cite{Engel2010} motion of the molecular gas, traced by the CO emission, rather aligns to the turbulence motion than to the circular movements. Recently obtained with MUSE instrument velocity dispersions correspond to N nucleus BH mass $(3.6\pm0.8)\times10^{8}\rm\;M_{\odot}$ and combined S (S1+S2) nucleus BH mass $(8.0\pm0.8)\times10^{8}\rm\;M_{\odot}$ \citep{Kollatschny2020}.

In the current work, we study the dynamical evolution of the SMBHB system in NGC~6240 using fully parallelised direct N-body code \PGPU ~\citep{Berczik2011}. This evolution has been examined by performing several simulations of the two SMBHs dynamics, each of which is surrounded by its own bound stellar systems. These simulations required the initial parameters of the binary nucleus in the NGC~6240, which were obtained from spectral analysis of archival \textit{Chandra} observations.

The paper is organised as follows. In Section~\ref{sec:XON}  we present the analysis of X-ray emission from nuclei and dynamical mass estimation. Working code and relativistic treatment of the binary particles are described in Section~\ref{sec:NM}. In Section~\ref{sec:InM} we describe a physical model and the set of numerical models for the NGC~6240 system based on our BH mass estimation. We applied our results to find the merging time for SMBHB and the expected GWs waveforms from this event in Section~\ref{sec:results}. Our conclusions are given in Section~\ref{sec:conc}. Throughout this paper we assume $\Lambda$CDM cosmology with a Hubble constant of $H_{0}=70$~km~s$^{-1}$~Mpc$^{-1}$, $\Omega_{\rm M}=0.27$ and $\Omega_{\Lambda}=0.73$ \citep{Bennett2003}, which gives a scale $1\arcsec=490$~pc \citep{Wright2006}. 

\section{\textit{Chandra} data analysis}\label{sec:XON}
\subsection{Image and spectral analysis}
\begin{figure*}
\includegraphics[width=0.32\linewidth]{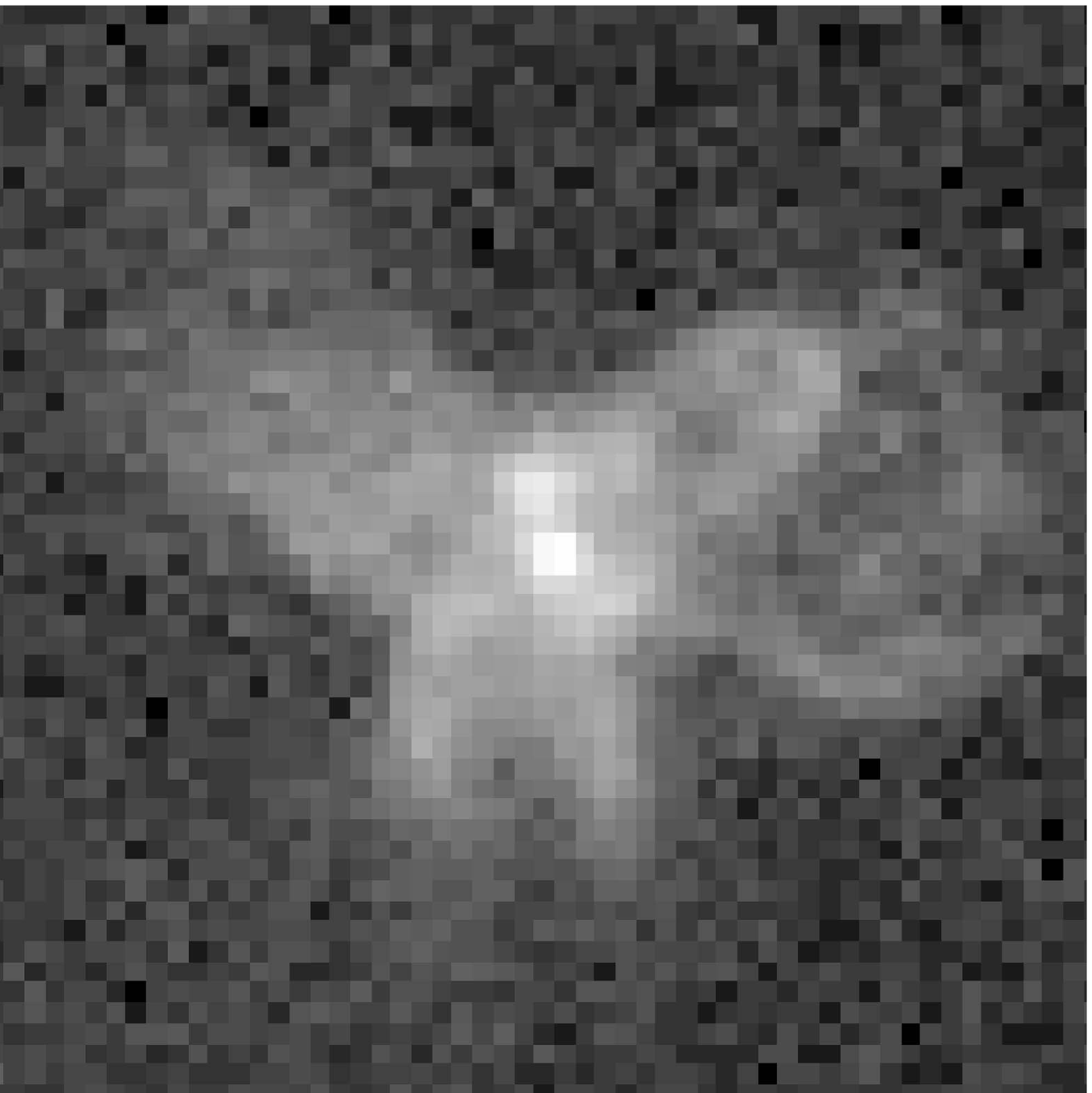}
\includegraphics[width=0.32\linewidth]{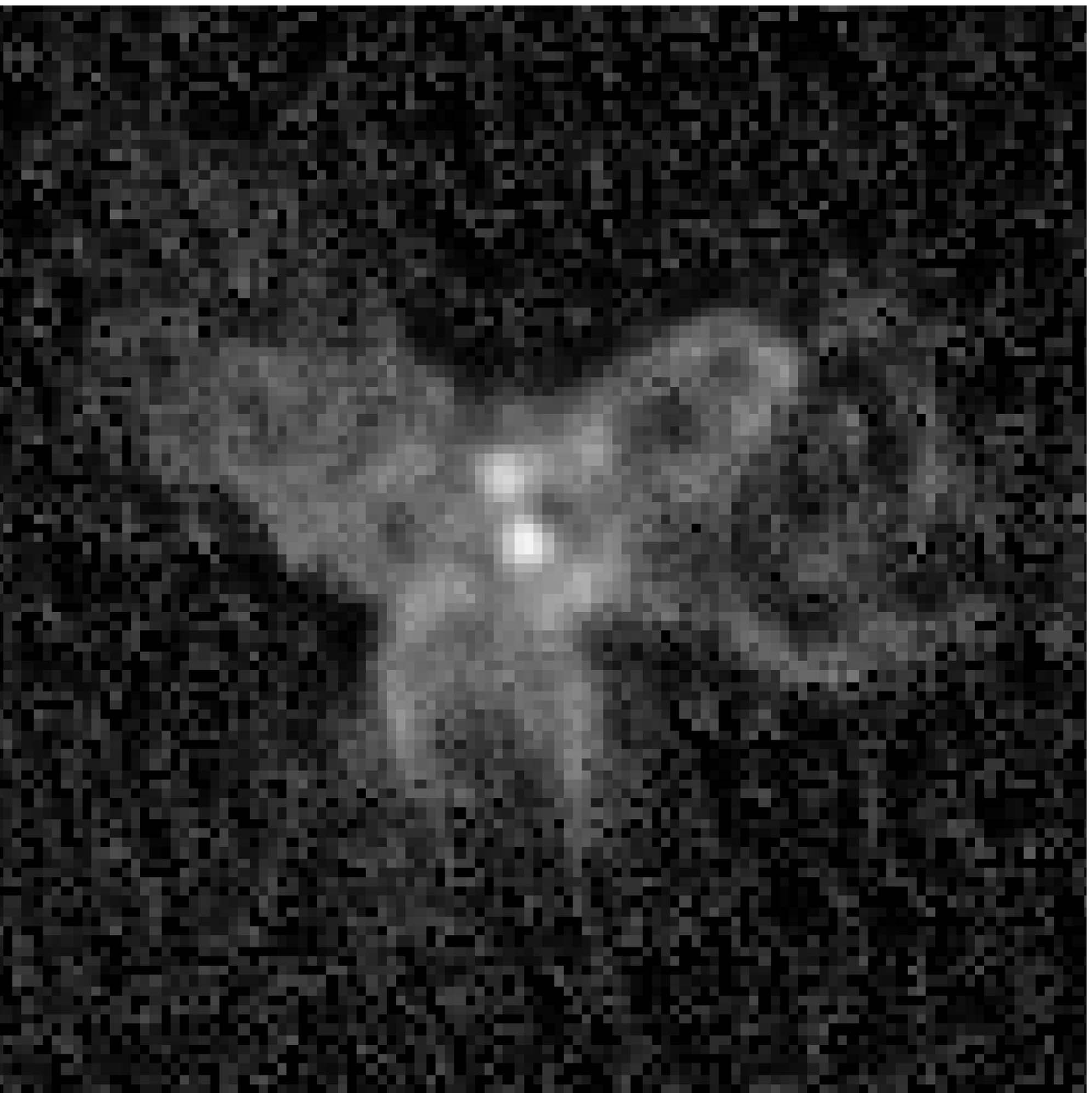}
\includegraphics[width=0.32\linewidth]{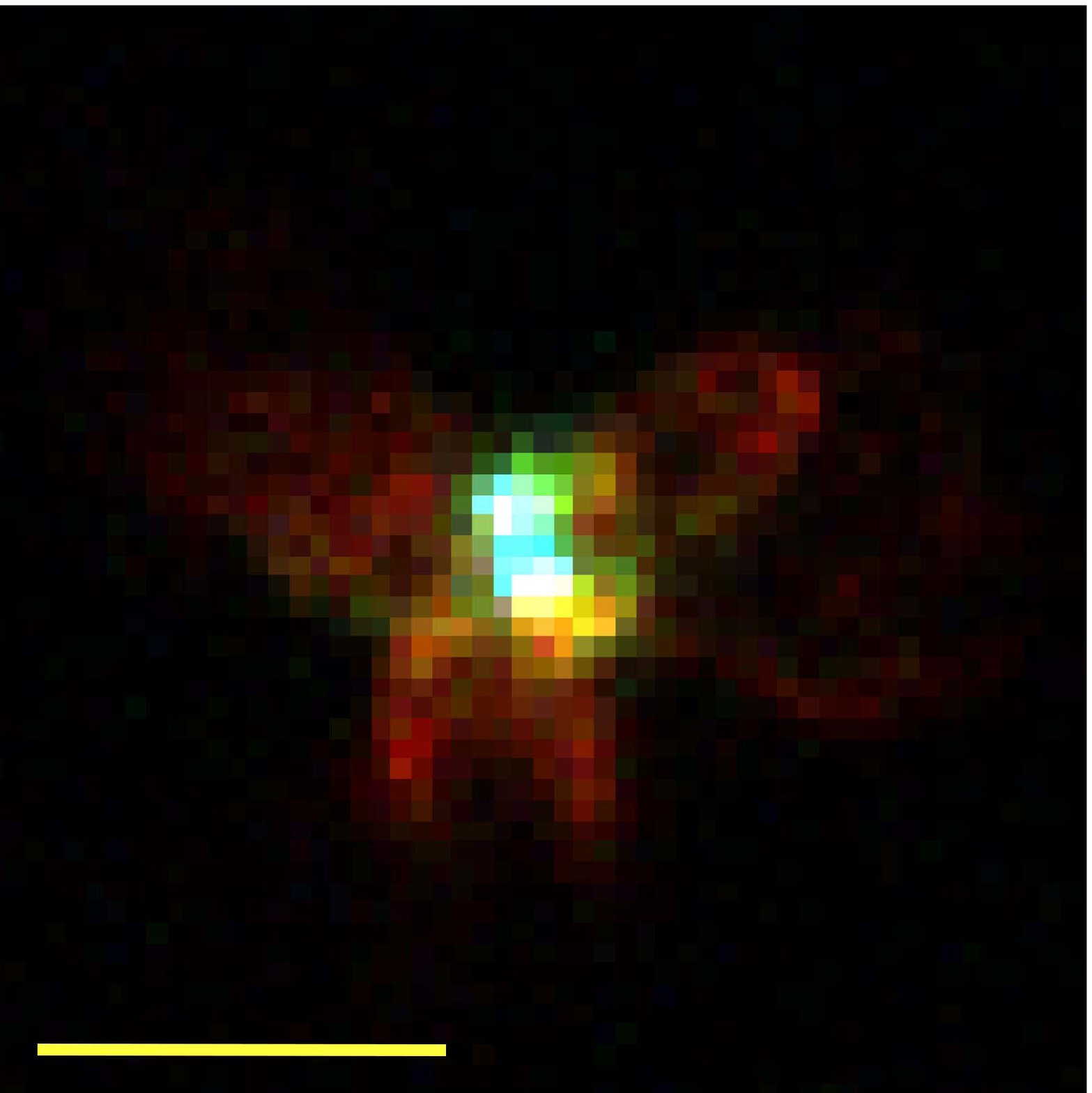}
\caption{The \textit{Chandra} images of NGC~6240 with the binning factor of 0.5: (left) the original merged image, (middle) the deconvolved image, (right) the exposure-corrected three-colour image, where colours correspond to energies: red:~0.5--2.5~keV, green:~2.5--6.0~keV, blue:~6.0--7.5~keV. The yellow line on the right panel is 5~kpc long.}
\label{fig:MergImg}
\end{figure*}

NGC~6240 was observed by \textit{Chandra} four times by Advanced CCD Imaging Spectrometer (ACIS) and once by High Resolution Camera (HRC). In the present work, we used only ACIS observations (ObsID 1590, 6908, 6909, 12713) with a total effective exposure time of 480.3~ks. The analysis of \textit{Chandra} data was done with the {\tt CIAO\,4.12} software package \citep{Fruscione2006} and the calibration database {\tt CALDB\ 4.9.1}. Before the analysis, the data were reprocessed using the {\tt chandra\_repro} script recommended in the {\tt CIAO} analysis threads.

Firstly, the \textit{Chandra} images in different energy bands (0.5--2.5~keV, 2.5--6.0~keV and 6.0--7.5~keV) were studied for carefully extracting the spectra from the regions corresponding to central BHs. We combined four ACIS observations using the {\tt merge\_obs} script from {\tt CIAO} software package and created the exposure-corrected image (Fig.~\ref{fig:MergImg}, right). It shows that the neutral Fe~K$\alpha$ emission lines were produced only in the central region of the galaxy which accords with the results presented by \cite{Komossa2003}.

We restored the image to analyse the detailed spatial structure since the original X-ray data are degraded by the blurring function. To restore the image we applied the Lucy--Richardson Deconvolution Algorithm (LRDA) implemented in the {\tt CIAO} tool {\tt arestore}. This algorithm requires an image of Point Spread Function (PSF) which was modelled by the {\tt ChaRT} and {\tt MARX} programs for detailed ray-trace simulations \citep{Carter2003,Wise1997} (Fig.~\ref{fig:MergImg}, left and middle). Consequently, we simulated the PSF for energy $E=6.25$~keV since we were interested mostly in the analysis of the central part of the galaxy where the emission is dominated by Fe~K$\alpha$ line. Two separate nuclei are more clearly visible due to the deconvolution (Fig.~\ref{fig:MergImg}, middle). Furthermore, the galaxy butterfly-shape in X-ray band matches optical with \ion{O}{III} cone, H$\alpha$ bubble, H$\alpha$ filaments and \ion{O}{III}+H$\alpha$ filaments, which are a consequence of galaxies merging history \citep{Muller2018}.

The spectra were extracted from circular regions centred at the centroid position of two bright sources in the galaxy nuclei. The each radius was determined as  3$\sigma$ encircled count fraction regions of the correspondent PSFs that were separately modelled for the S and N nuclei. The sum of these regions' diameters is 2\arcsec ($\approx1$~kpc) and can be taken as the maximum separation between the nuclei.

\begin{table}
\caption{The best-fitting parameters for X-ray spectra from Northern~(N) and Southern~(S) nuclei.}
\label{table:Chandra}
\centering
\renewcommand{\arraystretch}{1.3}
\begin{tabular}{l l l l}
\hline
\hline
Parameter & N & S & Unit \\
\hline
\hline
Galactic absorption                     & 0.0626$^{f}$             &  0.0626$^{f}$               & 10$^{22}$~cm$^{-2}$ \\
Photon index $\Gamma$                   & 1.75$^{f}$               & 1.75$^{f}$                  &  \\
Absorbing column density N$_{\rm H}$    & 5.00$^{+0.23}_{\rm-peg}$ & 31.30$_{-2.70}^{+2.40}$     & 10$^{22}$~cm$^{-2}$   \\
Line centre energy Fe~K$\alpha$   	& 6.41$_{-0.02}^{+0.01}$   & 6.39$_{-0.02}^{+0.01}$      & keV  \\
Line width $\sigma_{\rm Fe~K\alpha}$ & 0.05$^{+0.01}_{-0.02}$   & 0.05$_{-0.03}^{+0.04}$      & keV  \\ 
Line centre energy \ion{Fe}{XXV}       	& 6.72$_{-0.05}^{+0.06}$   & 6.66$_{-0.07}^{+0.06}$      & keV  \\
Line width $\sigma_{\ion{Fe}{XXV}}$     & 0.01$^{+0.12}_{\rm-peg}$ & 0.04$^{+0.22}_{\rm-peg}$    & keV  \\
Line centre energy Fe~K$\beta$       & 7.02$_{-0.04}^{+0.07}$   & 7.00$_{-0.21}^{+0.05}$      & keV  \\
Line width $\sigma_{\rm Fe~K\beta}$  & 0.09$^{+0.18}_{-0.08}$   & 0.04$^{\rm+peg}_{\rm-peg}$  & keV  \\
Reduced ${\rm\chi ^{2}/d.o.f}$ 	        & 179.5/175 		  	   & 164.1/196		             & 	    \\
\hline
\end{tabular}
\begin{minipage}{\linewidth}
\smallskip
NOTE:
\textit{f} -- marks a fixed parameters,
peg -- indicates a zero error,
d.o.f -- degrees of freedom.
\end{minipage}
\end{table}

For the spectral analysis, we extracted the corresponding spectra from each ObsID using the {\tt specextract} tool from the {\tt CIAO} software package. The background spectrum was created for a circular region located outside the galaxy and subtracted from nuclei spectra. To take into account the telescope response we created the Auxiliary Response Files (ARF) and the Redistribution Matrix Files (RMF) separately for each ObsID. The spectra of the four ACIS observations for each region were combined using the {\tt combine\_spectra} script from {\tt CIAO} software package. The data were grouped by {\tt group\_snr()} to set the minimum value signal-to-noise ratio for each bins and fitted using Sherpa \citep{Freeman2001} fitting application.

\begin{figure*}
\centering
\includegraphics[width=0.48\textwidth]{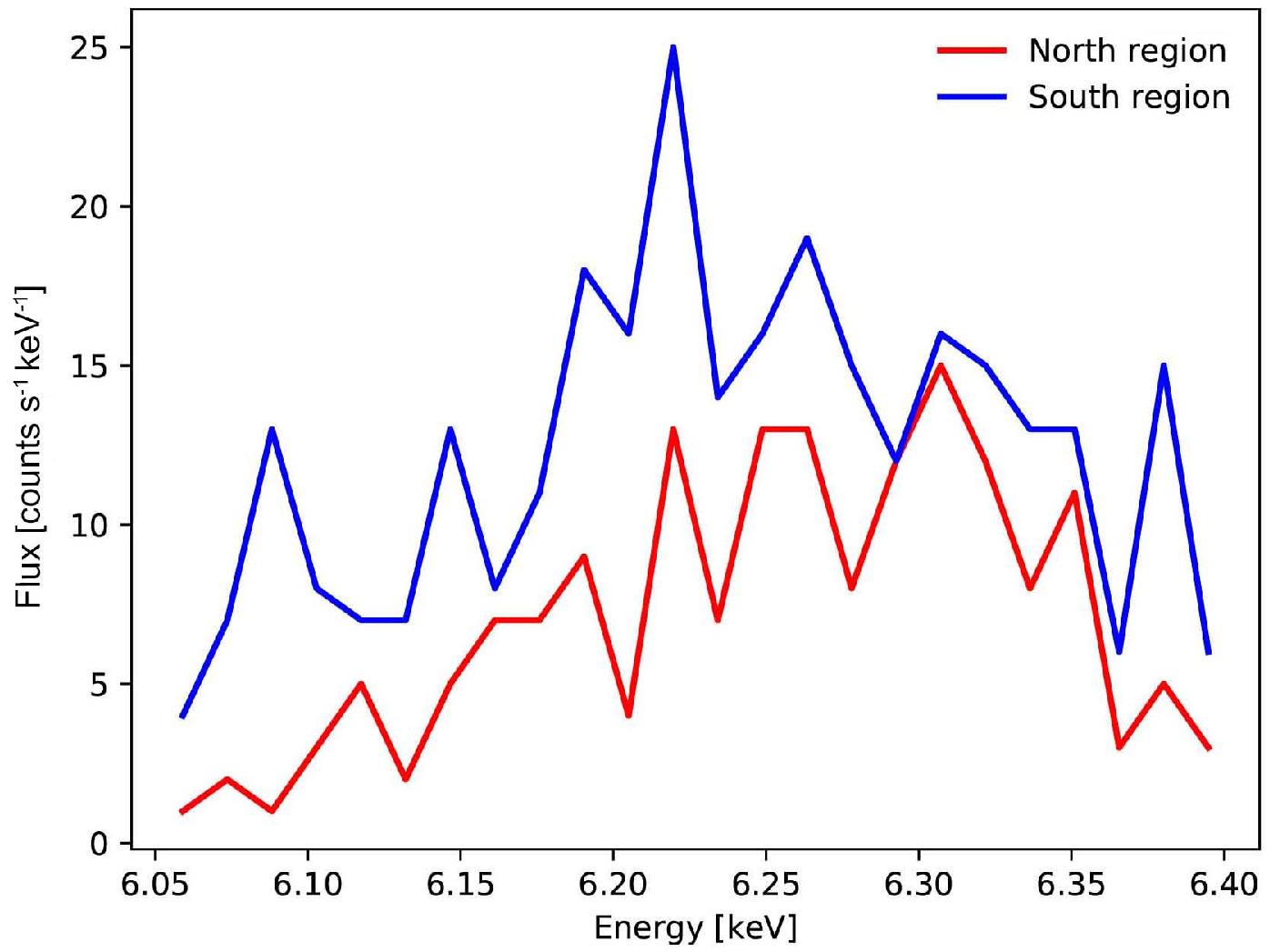}
\includegraphics[width=0.48\textwidth]{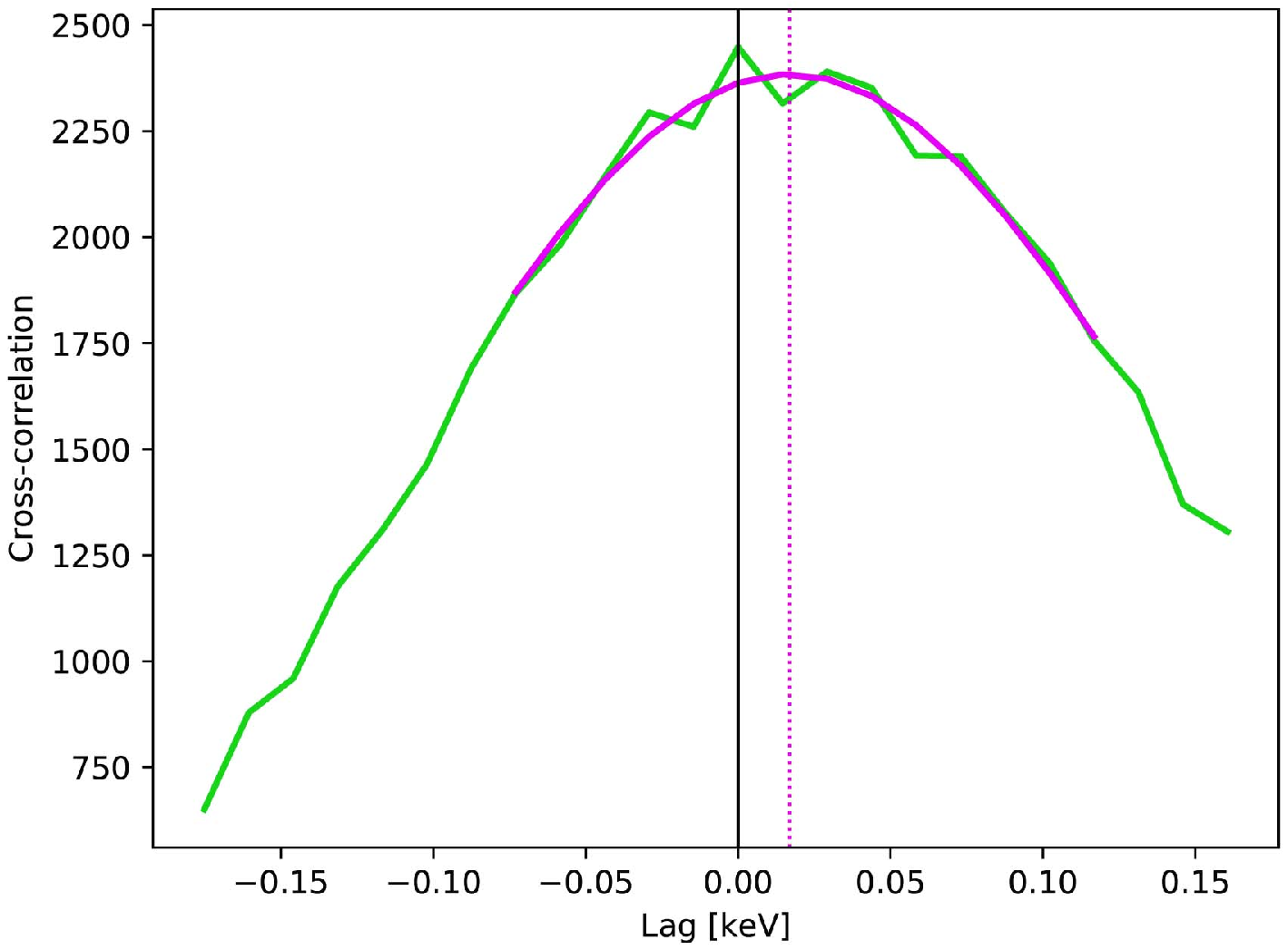}
\caption{The original X-ray spectra for N and S nuclei (left) and their cross-correlation (right -- green line). The magenta curve on the right panel corresponds to a best-fitting Gaussian, where the position of its centre is shown by the magenta dashed line.}
\label{fig:CC}
\end{figure*}

The spectra were described in energy range 5.5--7~keV using simple phenomenological model that includes the power law ({\tt xszpowerlw}), Galactic absorption ({\tt xsphabs}), and absorption on the line-of-sight ({\tt xszphabs}). We also added the Gaussian profiles ({\tt xszgauss}) for model of the neutral Fe~K$\alpha$ fluorescent emission line at 6.4~keV, He-like iron \ion{Fe}{XXV}~K$\alpha$ emission line at 6.7~keV and Fe~K$\beta$ fluorescent emission line at 7.08~keV. The photon indices were fixed for nuclei with value~ $\Gamma = 1.75$. Detailed broadband analysis for continuum spectrum is presented in \cite{Puccetti2016} and \cite{Nardini2017}. Finally the model in Sherpa was described as follows:
\begin{center}
\texttt{xsphabs $\ast$ (xszpowerlw $\ast$ xszphabs + xszgauss + xszgauss + xszgauss)}
\end{center}
We compiled the best-fitting parameters in Table~\ref{table:Chandra}. The Fe~K$\alpha$ best-fit values are $E_{\rm N} = 6.41_{-0.02}^{+0.01}$~keV and  $E_{\rm S} = 6.39_{-0.02}^{+0.01}$~keV for N and S nuclei respectively. Therefore, line shift $\Delta E \approx0.02$~keV can be interpreted as the result of the motion of each nucleus around the centre of mass. The Fe~K$_\alpha$ emission lines widths are $\sigma_{\rm N} = 0.05^{+0.01}_{-0.02}$~keV and $\sigma_{\rm S} = 0.05_{-0.03}^{+0.04}$~keV for N and S nuclei respectively. Such values of emission lines widths mean that the Fe~K$\alpha$ line is a narrow one. The emission lines at $6.72^{+0.06}_{-0.05}$ and $6.66^{+0.06}_{-0.07}$~keV can be explained as a highly ionised \ion{Fe}{XXV} emission from circumnuclear starburst regions \citep{Wang2014a}.

\subsection{Mass estimation}\label{sec:mass_est}
Assuming that N and S nuclei have formed the bound system and move around the mass centre we can estimate the enclosed mass. Based on the energy shift ${\Delta E}$ of the observed Fe~K$\alpha$ lines and their line centre mean energy $E_{\rm obs} = 0.5(E_{\rm N}+E_{\rm S})$ from Table~\ref{table:Chandra} we obtained the velocity shift:
\begin{equation}
\Delta \varv_{\rm obs} = \frac{\Delta E c}{E_{\rm obs}} \approx 937\rm\;km\;s^{-1},
\label{eq:v_obs}
\end{equation}
where $\Delta E$ is the energy shift between two nuclei. We collected velocity differences from other bands in Table~\ref{tab:vel} and found that in comparison with optical/IR and radio observations our $\Delta\varv_{\rm obs}$ is a factor of three higher. Should be mentioned that this comparison is restricted by several limitations: (i) in most cases values were obtained after simple visual inspection of velocity maps, which also limited us in velocity error estimations; (ii) different AGNs’ coordinates were used in observations, which complicated maps matching in different bands; (iii) the chosen of the region for velocity extraction is unclear, also are complicated by resolutions in different bands (from 0\farcs5 in X-ray band to 0\farcs03 in mm band). We assume that this discrepancy can be explained with the model where the X-ray and optic/radio bands emission is created in physically different regions at significantly different distances from the central BH. 

\begin{table}
\caption{Absolute velocity difference between nuclear regions from X-ray, radio and optic/IR bands.}
\centering
\renewcommand{\arraystretch}{1.2}
\sisetup{separate-uncertainty}
\resizebox{0.47\textwidth}{!}{
\begin{tabular}{l@{\hspace{0.9\tabcolsep}}l@{\hspace{0.9\tabcolsep}}l@{\hspace{0.9\tabcolsep}}
S[table-format=3.0(2)]@{\hspace{0.9\tabcolsep}}
S[table-format=1.2]@{\hspace{0.9\tabcolsep}}
S[table-format=3.0]@{\hspace{0.9\tabcolsep}}
c}
\hline
\hline
Band & Instrument & Line & {$\Delta \varv$} & \multicolumn{2}{c}{{\hspace{-1.0em}Resolution}} & Reference\hyperlink{link3}{$^\dag$} \\
     &            &      & {km s$^{-1}$}    & {\arcsec}   & pc         &           \\ 
\hline
\hline
X-ray & \textit{Chandra} & Fe~K$\alpha$                       & 937      & 0.5  & 245 & This paper \\
NIR   & SINFONI          & CO(2-0)+CO(3-1)                    & 252\pm15 & 0.1  &  49 & [1] \\
NIR   & SINFONI          & H$_{2}$                            & 250      & 0.5  & 245 & [2] \\
NIR   & SINFONI          & [\ion{O}{III}] $\lambda$5007       & 150      & 0.5  & 245 & [2] \\
IR    & MUSE             & \ion{Ca}{II} $\lambda\lambda$8498, & 144\pm42 & 0.03 &  15 & [3] \\
      &                  & \qquad\quad\;\;8542, 8662 & & & & \\
IR    & MUSE             & [\ion{N}{II}] $\lambda$6548        & 160\pm54 & 0.03 &  15 & [3] \\
IR    & MUSE             & [\ion{O}{I}] $\lambda$6300         & 262\pm24 & 0.03 &  15 & [3] \\
Radio & ALMA             & CO(3-2)                            & 250      & 0.3  & 147 & [4] \\
Radio & ALMA             & CO(6-5)                            & 100      & 0.3  & 147 & [4] \\
Radio & ALMA             & $^{12}$CO(2-1)                     & 300      & 0.03 &  15 & [5] \\
Radio & IRAM             & H$_{2}$                            & 200      & 0.1  &  49 & [1] \\
\hline
\end{tabular}}
\begin{minipage}{\linewidth}
\smallskip
\hypertarget{link3} $^\dag$References:
[1]~\cite{Engel2010},
[2]~\cite{Muller2018},
[3]~\cite{Kollatschny2020},
[4]~\cite{Fyhrie2021} 
[5]~\cite{Treister2020}.
\end{minipage}
\label{tab:vel}
\end{table} 

In interacting galaxy NGC~6240 we expected that the emission in Fe~K$\alpha$ line created in a pc scale (inside the gas-dusty torus; see e.g. \citealt{Nandra2006}) in contrast with the observed optic/IR emission which comes from the distance of tens of pc from the central SMBH. Recent studies of bright nearby AGN ($z < 0.5$) with \textit{Chandra} and \textit{XMM-Newton} data are showed that with high probability for 24 objects the narrow Fe~K$\alpha$ line is emitting from the inner 1 pc around SMBH \citep{Andonie2022}. The next generation of planned space-born X-ray observatories includes \textit{Athena} proposed by ESA \citep{Nandra2013,Barret2018} and \textit{Lynx} proposed by NASA \citep{Lynx2018,Gaskin2019}. They are expected to have a higher $\sim100$ times spectral resolution on 6~KeV, which can make clear the nature of the observed Fe~K$\alpha$ line.

The dynamical mass can be written in terms of observed velocity shift:
\begin{equation}
M_{\rm dyn} \approx \frac{\Delta R \Delta \varv_{\rm obs}^{2}}{G},
\end{equation}
where $\Delta R$ is the separation and $G$ is the gravitational constant. Using maximum projected distance $\Delta R_{\rm proj}=1$~kpc as a estimation for the minimum physical separation of SMBHB $\Delta R=\Delta R_{\rm proj}$ we obtained the total dynamical mass within this region $M_{\rm dyn} \approx 2.04\times 10^{11}{\rm M}_{\odot}$. Of course, our dynamical mass estimation affected by the underlying assumptions about the simple geometry of the NGC~6240 central region. As a first approximation, we assume that the projected separation of the nuclei is a intrinsic size of the system. We also assume that the observed velocity shift between nuclei is a real velocity difference. The current simple assumptions we use as a basis for our BHs dynamical merging time estimation at a first order. The detail parameter study of the possible different orientation and projection of the nuclei we keep beyond the scope of the current paper.

According to the empirical correlation between SMBH and galaxy bulge masses \citep{Kormendy2013}, and due to active merging galaxy state we estimated the maximum SMBHB total mass $M_{\rm BH12}=0.01M_{\rm dyn} \approx 2.04\times10^9\rm\;M_{\odot}$. The obtained mean mass $M_{\rm BH}$ is comparable with the dynamical masses previously derived by \cite{Medling2011} and \cite{Kollatschny2020}.

The difference $\Delta E$ between the Fe~K$_\alpha$ lines centroids in spectra of both nuclei is the same order as the errors of two line's positions. Therefore, we performed additional validation of the estimated difference $\Delta E$, using the cross-correlation between N and S nuclei spectra (Fig.~\ref{fig:CC}, left). The cross-correlation between original spectra is presented in Fig.~\ref{fig:CC}~(right), where the magenta line is the fitted Gaussian function. The best-fitting position of the Gaussian profile is 0.0170$\pm$0.0019~keV, which is consistent with the estimated shift that we got from spectral fitting within the errors.

\section{Numerical modelling SMBH particles}\label{sec:NM}
For our simulations we used our own developed and publicity available \PGPU{\footnote{\hypertarget{link1}{\url{https://github.com/berczik/phi-GPU-mole}}}} code with the blocked hierarchical individual time step scheme and a $4^{\rm th}$-order Hermite integration scheme of the equation of motions for all particles \citep{Berczik2011, 2008AN....329..904B}. The current version \PGPU~code uses a native GPU support and direct code access to the GPU using the NVIDIA CUDA library. The multi GPU support is achieved through the global MPI parallelisation. Each MPI process uses only a single GPU, but usually up to four MPI processes per node are started (in order to effectively use the multi core CPUs and the multiple GPUs on our clusters). More details about the $\varphi$-GPU code public version and its performance we are presented in 
\cite{2018A&A...615A..71K} and \cite{2012MNRAS.419...57F}. The present code is well tested and already used to obtain important results in our earlier large scale (up to few million body) simulations \citep{2014ApJ...792..137Z, 2018ApJ...868...97K, 2012ApJ...748...65L, 2014ApJ...780..164W}. For simulations with lowest particle number $N=100$k we used the {\tt GOLOWOOD} GPU cluster at MAO NASU. The main part of our numerical experiments with the largest particle number ($N = 500$k and 200k) we run on the {\tt JUWELS} GPU cluster of the J\"ulich Supercomputing Centre.

In the current implementation of the code, we used a post-Newtonian ($\mathcal{PN}$) formalism for the SMBHB relativistic orbit calculation. In this case, the equation of motion is usually presented as a power series $1/c$ of light velocity, where n-$\mathcal{PN}$ is proportional to $(\varv/c)^{2\rm n}$. The acceleration of the $i$ binary particle from $j$ particle with mass $m_{j}$ one can write in the following form:
\begin{equation}\label{eq:schem}
\boldsymbol{a}_{i} = -\frac{Gm_{j}}{R^{2}}[(1+\mathcal{A}\boldsymbol{n}_{ij})+\mathcal{B}\boldsymbol{v}_{ij}],
\end{equation}
where $R$ is the separation between $i$ and $j$ binary particles, $\boldsymbol{n}_{ij}$ is the normalised position vector and $\boldsymbol{v}_{ij}$ is the relative velocity vector. The classic Newtonian acceleration has explicit representation in equation~(\ref{eq:schem}), when $\mathcal{PN}$ corrections are contained in two coefficients $\mathcal{A}$ and $\mathcal{B}$:
\begin{equation}
\mathcal{A}=\frac{\mathcal{A}_{1\mathcal{PN}}}{\rm c^{2}}+
\frac{\mathcal{A}_{2\mathcal{PN}}}{\rm c^{4}}+
\frac{\mathcal{A}_{2.5\mathcal{PN}}}{\rm c^{5}}+
\mathcal{O}\Bigr(\frac{1}{\rm c^{6}}\Bigl),
\end{equation}
\begin{equation}
\mathcal{B}=\frac{\mathcal{B}_{1\mathcal{PN}}}{\rm c^{2}}+
\frac{\mathcal{B}_{2\mathcal{PN}}}{\rm c^{4}}+
\frac{\mathcal{B}_{2.5\mathcal{PN}}}{\rm c^{5}}+
\mathcal{O}\Bigr(\frac{1}{\rm c^{6}}\Bigl),
\end{equation}
where $1\mathcal{PN}$ and $2\mathcal{PN}$ are the non-dissipative terms that `conserve' the energy of the system and are revealed in the precession of the orbital pericenter. The $2.5\mathcal{PN}$ is the dissipative term that `carries out' energy from the system due to GWs emission. Coefficients $\mathcal{A}$ and $\mathcal{B}$ are the functions of individual masses, individual velocities, separation and normalised vector. Their full expressions can be found in \citet[][equation 168]{Blanchet2006}. The detailed references and complete descriptions of the equation of motion in $\mathcal{PN}$ formalism up to $3.5\mathcal{PN}$ can be found at \cite{Blanchet2006}, \cite{Kupi2006}, \cite{2008AN....329..904B}, \cite{Berentzen2009}, \cite{Brem2013}, \cite{Sobolenko2017}.

Detailed study of the turning on one-by-one $\mathcal{PN}$ corrections show requirement to including all $\mathcal{PN}$ terms up to highest wanted order \citep{Berentzen2009}. Adding conservative $1\mathcal{PN}$ and $2\mathcal{PN}$ corrections remarkably change orbits during three-body encounters and can reduce binary merging time two times. We applied all $\mathcal{PN}$ corrections up to order~$\mathcal{O}(1/c^{6})$, so the $2.5\mathcal{PN}$ correction is the highest order that we took into account.

\section{System initial configurations}\label{sec:InM}
\subsection{Physical model and units}\label{sec:phys_m}
\begin{table}
\caption{List of parameters for physical model.}
\label{tab:phys_models}
\centering
\renewcommand{\arraystretch}{1.2}
\begin{tabular}{ccrccrc}
\hline
\hline
Nucleus & $\Delta R$ & \multicolumn{1}{c}{$M_{\ast}$} & $Q$ & $a$ & \multicolumn{1}{c}{$M_{\rm BH}$} & $q$ \\
 & kpc & $10^{10}\rm\;M_{\odot}$ & & pc & $10^{8}\rm\;M_{\odot}$ & \\
(1) & (2) & \multicolumn{1}{c}{(3)} & (4) & \multicolumn{1}{c}{(5)} & \multicolumn{1}{c}{(6)} & (7) \\
\hline
\hline
1~(N) & 1  & 13.60 & 0.5 & 200 & 13.60 & 0.5 \\
2~(S) & -- &  6.80 & --  & 159 &  6.80 & -- \\
\hline
\end{tabular}
\begin{minipage}{\linewidth}
\smallskip
NOTE:
(1) nuclei ID,
(2) initial separation for central BHs,
(3) total stellar mass,
(4) stellar mass ratio $Q=M_{\ast2}/M_{\ast1}$,
(5) Plummer radius,
(6) masses of the BHs,
(7) mass ratio for the BHs $q=M_{\rm BH2}/M_{\rm BH1}$.
\end{minipage}
\end{table}

The evolution of the central parts of the merging galaxies is closely related to the dynamical processes of the SMBHB evolution. The stars located in the merging galactic centre can interact directly with the SMBHB. Such stars in close orbits around the SMBHB can take away a significant part of the SMBHB angular momentum and energy after the typical three-body gravitational scattering. As a result the semimajor axis of the binary system monotonically decreases. This process we usually call SMBHB `hardening' \citep{Merritt2001a, Merritt2001b}. Very precise individual orbit calculation of the merging SMBHB in a dense stellar environment gives the correct description of binary system parameters evolution.

We started the galaxy merger from the dynamical system of two unbound central SMBHs with the separation $\Delta R=1$~kpc according to our estimations in Section~\ref{sec:mass_est} (Table~\ref{tab:phys_models}). Each SMBH is surrounded by its own bound stellar systems with simple Plummer density distribution \citep{Plummer1911}:
\begin{equation}
\rho(r) = \frac{3 M_0}{4 \pi a^3} \left( 1+\frac{r^2}{a^2} \right)^{-\frac{5}{2}},
\end{equation}
which produce the cumulative mass distribution:
\begin{equation}
M(< r) = M_0 \frac{r^3}{(r^2 + a^2)^{3/2}},
\end{equation}
where $M_0$ is the total mass of each galactic bulge, and $a$ is a scale factor that characterises the size of each nucleus (Plummer radius). Due to the flat central distribution of the Plummer profile, the SMBHB hardening as the assuming numerical hardening will be smaller compared to the more peaked core distribution profiles \citep{Jaffe1983,Hernquist1990,Dehnen1993}. Using Plummer distribution we model the minimum numerical hardening for our SMBHB.

Previously estimated from observations dynamical mass is assumed as the total mass of the stellar component $M_{\ast,\rm tot}=M_{\rm dyn} = 2.04\times10^{11}\rm\;M_{\odot}$. Corresponding to Section~\ref{sec:mass_est} the mass of the SMBHB is set $M_{\rm BH12}=2.04\times10^{9}\rm\;M_{\odot}$. Supposing the major merging we used for the mass ratio of the galactic bulges and the central BH's 2:1 ratio. According to this assumption the primary (heavier) bulge with mass $M_{\ast1}=1.36\times10^{11}\rm\;M_{\odot}$ 
contains BH with mass 
$M_{\rm BH1}=1.36\times10^{9}\rm\;M_{\odot}$ 
and secondary (lighter) bulge with mass 
$M_{\ast2}=6.8\times10^{10}\rm\;M_{\odot}$ 
contains BH with mass 
$M_{\rm BH2}=6.8\times10^{8}\rm\;M_{\odot}$ 
(Table~\ref{tab:phys_models}). Also for further reference we calculated the Schwarzschild radius of the SMBHB as
${R_{\rm SW12}}=2GM_{\rm BH12}/c^{2}=195\rm\;\upmu pc$.

For the first bulge we assumed the Plummer radius near equal to the influence radius of the BH, that give 
$a_{1}=0.2\Delta R=200$~pc. 
For the second (smaller) bulge we set the Plummer radius proportionally smaller, assuming the same central density in both bulges, that is
$a_{2}=0.5^{1/3}a_{1}\approx159$~pc (Table~\ref{tab:phys_models}). The initial orbital velocities of the merging galactic bulges (together with the BH's) we set as that the orbital eccentricity (in point mass approximation) equals to $ecc_{0}=0.5$.

For the numerical scaling we used the N-body normalisation \citep{Henon1971}\footnote{\url{https://en.wikipedia.org/wiki/N-body\_units}}. The physical units was choosing according to estimations for total stellar mass and maximum projected separation between BHs:
\begin{equation}
{\rm M_{NB}} = M_{\rm dyn} = 2.04 \times 10^{11}~{\rm M}_{\odot},
\end{equation}
\vspace{-2em}
\begin{equation}
{\rm R_{NB}} = \Delta R = 1 {\rm~kpc}.
\end{equation}
In N-body system of units we have for velocity and  time units the rescaling values:
\begin{equation}
{\rm V_{NB}} = 936.7~{\rm km~s^{-1}},
\end{equation}
\vspace{-2em}
\begin{equation}
{\rm T_{NB}} = 1.04 {\rm~Myr}.
\end{equation}
In this system of units \citep{Sobolenko2017} for the light speed  we got the value: $c=320\rm\;V_{NB}$. 

\subsection{Numerical models}\label{sec:num_m}
\begin{table}
\caption{List of parameters for basic and mass prescription numerical models.}
\label{tab:num_models}
\centering
\renewcommand{\arraystretch}{1.2}
\begin{tabular}{ccrrrc}
\hline
\hline
$N$ & RAND & $m_{\rm HMP}$:$m_{\rm LMP}$ & $m_{\rm HMP}$ & $m_{\rm LMP}$ & $\mathcal{PN}$ \\
 & & & $10^{6}\rm\;M_{\odot}$ & $10^{6}\rm\;M_{\odot}$ & \\
(1) & (2) & \multicolumn{1}{c}{(3)} & \multicolumn{1}{c}{(4)} & \multicolumn{1}{c}{(5)} & (6) \\
\hline
\hline
100k &  1, 2, 3	                    &  10:1  & 10.20 & 1.130 & 1 \\
200k &  1, 2, 3                     &  10:1  &  5.10 & 0.567 & 2 \\
500k &  1, 2, 3	                    &  10:1  &  2.04 & 0.227 & 3 \\
\hline
100k &  1                           &  1:1   &   --  & 2.400 & --  \\
100k &  1                           &  5:1   &  5.10 & 1.280 & --  \\
100k &  1                           &  20:1  & 20.40 & 1.070 & --  \\
\hline
\end{tabular}
\begin{minipage}{\linewidth}
\smallskip
NOTE:
(1) total number of particles,
(2) randomisation seed number,
(3) HMPs to LMPs mass ratio,
(4) and (5) mass of HMPs and LMPs respectively,
(6) randomisation seed number for which we turned $\mathcal{PN}$ correction.
\end{minipage}
\end{table}

To check the numerical convergence of our Newtonian dynamical `hardening' timescale results we used three different total particle numbers for the system $N=100$k, 200k, and 500k. For each of these particle numbers, we run a separate set of simulations with three different particle random seeds RAND~=~1, 2, 3 (Table~\ref{tab:num_models}, top three numerical models). Below we will use the abbreviation 100-1 for a run with particle number $N=100$k and random seed RAND~=~1. In all of these nine basic runs, we generated two different types of particles for each galaxy (completely mixed inside the system), the so-called `high mass' (HMPs) and `low mass' particles (LMPs). We fixed the individual particles' mass ratio for these particles as 10:1. For all the nine runs we also used the fixed number ratio for the HMPs and LMPs particle number: $N_{\rm HMP}$:$N_{\rm LMP}$~=~1:10. This small fraction of HMPs allowed us to mimic the dynamical influence of the giant molecular clouds and/or the compact stellar systems (globular clusters) on the common stellar system of the merging centres (colliding bulges). Even these small fraction of `super' particles with a larger gravitational softening can have a great influence on the phase space mixing of the `normal' stellar particles.  

We also run three additional runs with $N=100$k simulations using the different HMPs to LMPs individual mass ratio. In comparison to the basic runs, where we set the ratio 10:1, we run simulations with mass ratios 5:1, 20:1 and with just LMPs without HMPs 1:1 (Table~\ref{tab:num_models}, bottom three models). We specially carried out these three runs to illustrate the dynamical effect of the possible higher mass ratio of the particles.

For different number of particles we also set a different individual gravitational softening length. For the BH-BH particles interaction we used the exactly zero softening ($\epsilon_{\rm BH} = 0.0$). For the HMPs we used $\epsilon_{\rm HMP} = 10^{-4}{\rm\;R_{NB}}=0.1$~pc gravitational softening. For the LMPs we set $\epsilon_{\rm LMP} = 10^{-5}{\rm\;R_{NB}}=0.01$~pc. For the mixed interactions between the different type of particles we used the mixed gravitational softening between the particles:
\begin{equation}
\epsilon^2_{ij} = 0.5 (\epsilon^2_{i} + \epsilon^2_{j}).
\label{eq:eps_ij}
\end{equation}
In a case, if one of the particles is a BH (or $i$ or $j$) we set the additional coefficient $10^{-2}$ in front of the equation~(\ref{eq:eps_ij}) to make a further extra reduction for such a gravitational interaction. As the result, we obtained effective softening parameters in level $10^{-5}{\rm\;R_{NB}}=0.01$~pc and $10^{-6}{\rm\;R_{NB}}=0.001$~pc for HMPs and LMPs respectively.

Leaned on the 9 basic Newtonian runs we run three full $\mathcal{PN}$ runs to leading SMBHB to merging, where we turned on the extra $\mathcal{PN}$ terms for the BH-BH gravitational interaction. Specially chosen three different Newtonian runs have different particle numbers and are noted with suffix $\mathcal{PN}$ (Table~\ref{tab:num_models}, top three numerical models). The $\mathcal{PN}$ terms turned on time $t_{\mathcal{PN}\rm beg}\approx10$~Myr after the binary binding at time $t_{\rm b}$ (Table~\ref{tab:times}). We stopped these runs when the SMBH particles separation fell below ~$\approx4\;R_{\rm SW12}$ and this time assumed as merging time $t_{\rm merge}$ (Table~\ref{tab:times}). We also will compare our results with previous simulations, which consist of 4 physical and 16 numerical models \citep[][hereafter \hyperlink{Sob21}{S21}]{Sobolenko2021}.

\begin{figure*}
\centering
\includegraphics[width=0.49\linewidth]{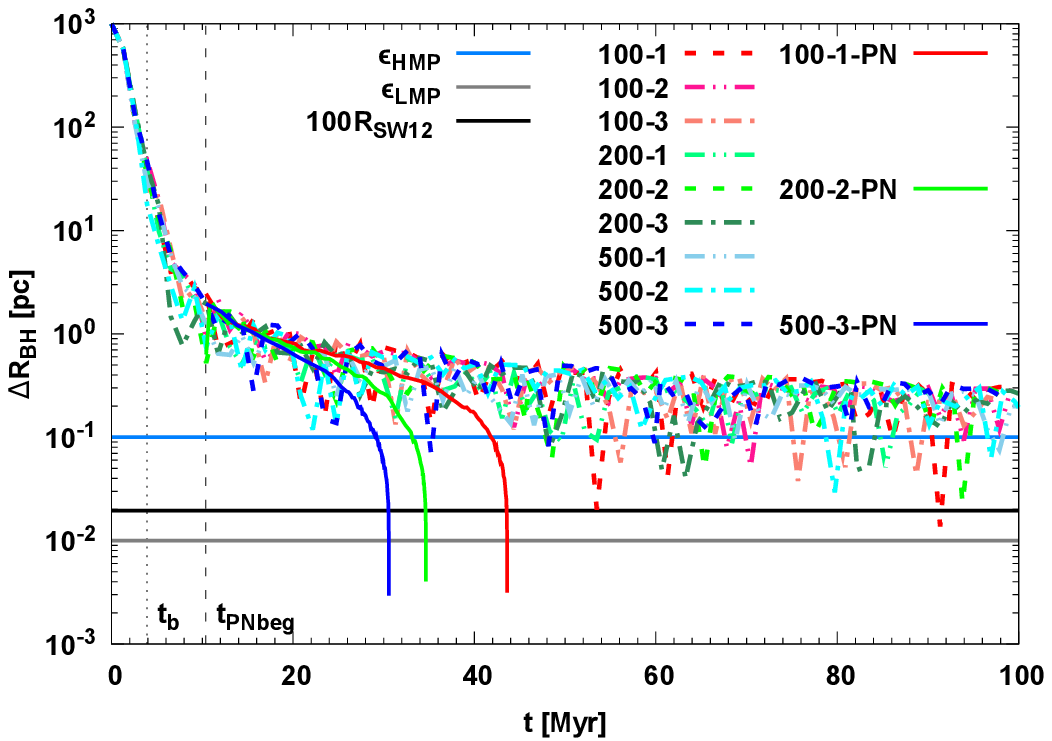}
\includegraphics[width=0.49\linewidth]{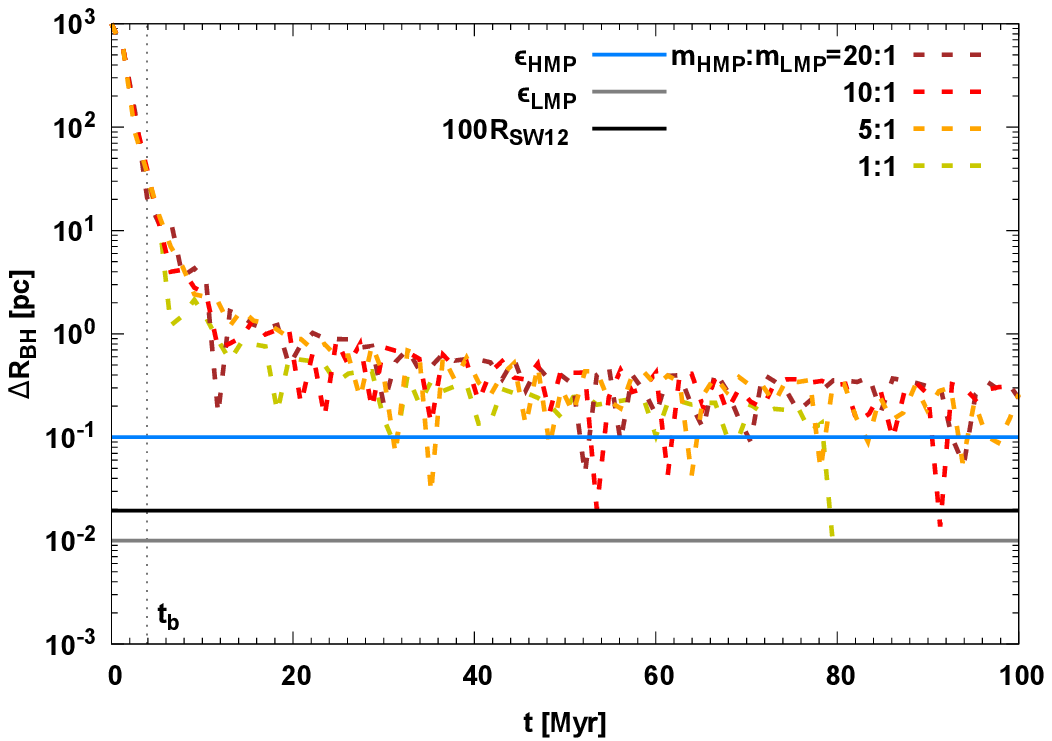}
\includegraphics[width=0.49\linewidth]{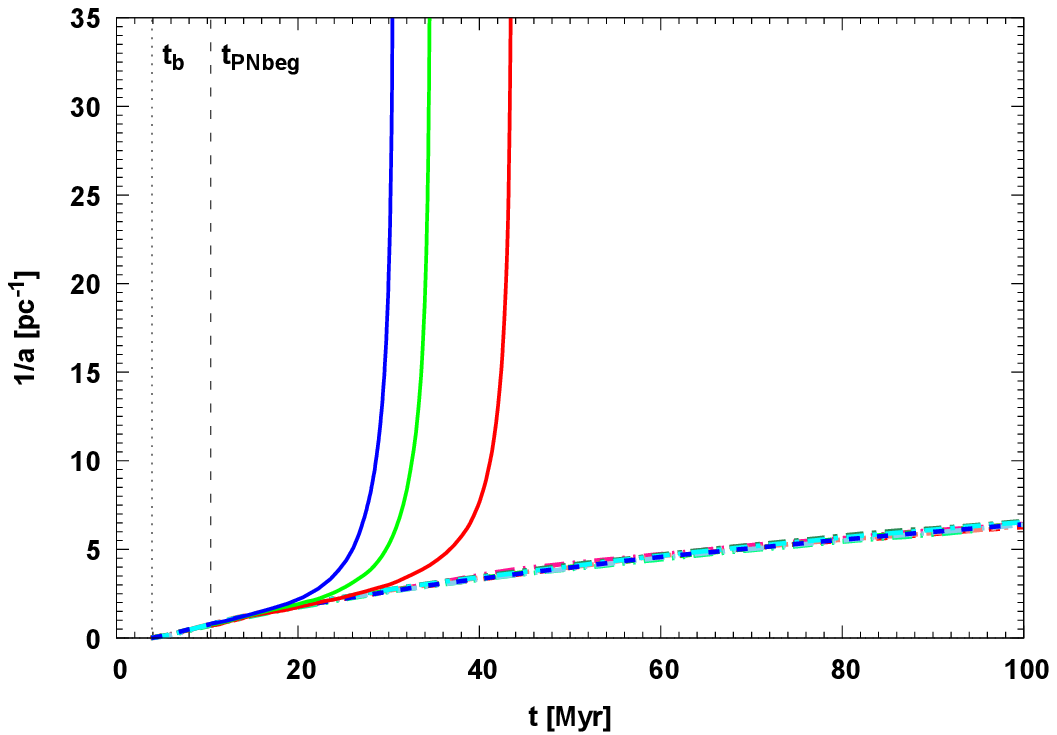}
\includegraphics[width=0.49\linewidth]{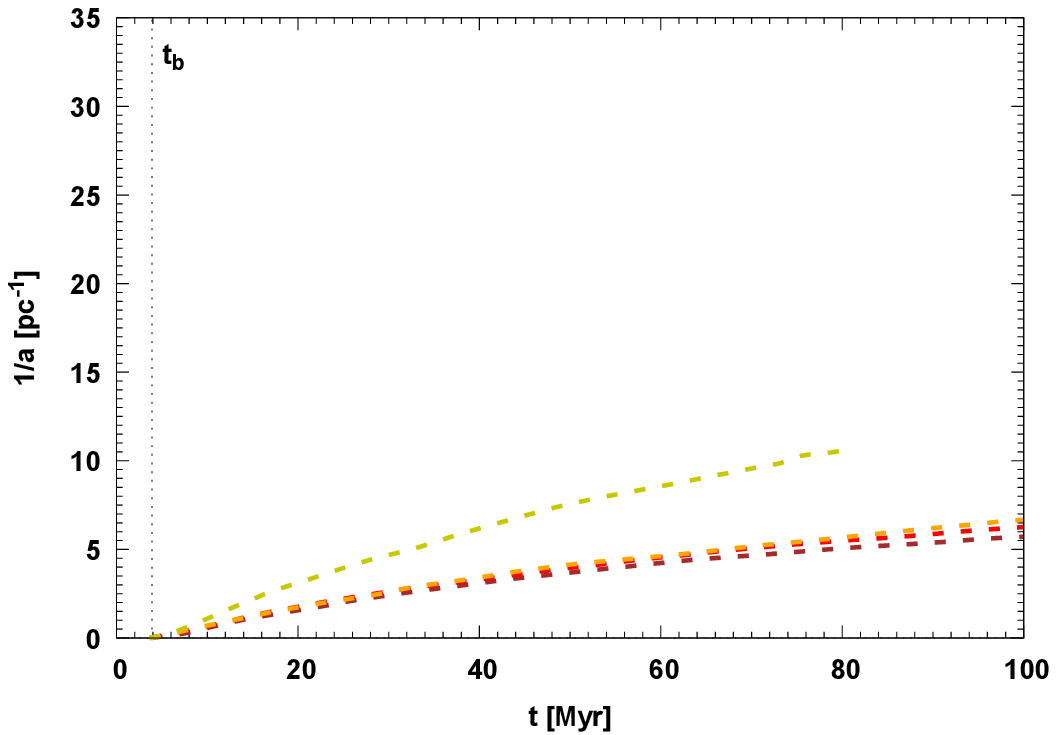}
\begin{minipage}{0.49\textwidth}
\includegraphics[width=\columnwidth]{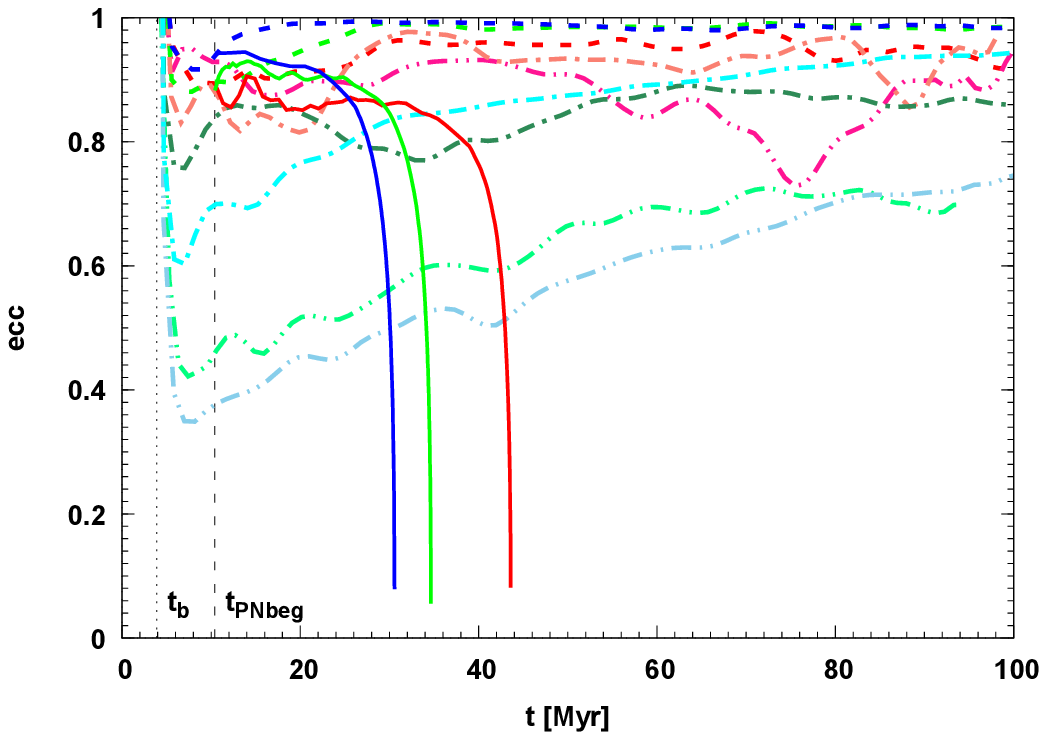}
\caption{Evolution of SMBHB separation (top), inverse semimajor axis (middle) and eccentricity (bottom) for basic Newtonian (colour dashed lines) and $\mathcal{PN}$ runs (colour solid lines) with mass ratios HMPs to LMPs 10:1 from Table~\ref{tab:num_models}. The red, green and blue solid lines are $\mathcal{PN}$ runs for particle number 100k, 200k, and 500k, respectively. On the top panel, the horizontal solid light blue and grey lines are softening parameters for HMPs and LMPs respectively, the solid black line is 100 Schwarzschild radii. Vertical black dashed lines are binding time $t_{\rm b}$ for models 100-1, 200-2, 500-3 (Table~\ref{tab:times}) with following turning on $\mathcal{PN}$ terms at time $t_{\mathcal{PN}\rm beg}$ (Table~\ref{tab:times}).}
\label{fig:models}
\end{minipage}\hfill
\begin{minipage}{0.49\textwidth}
\centering
\vspace{-6ex}
\includegraphics[width=\columnwidth]{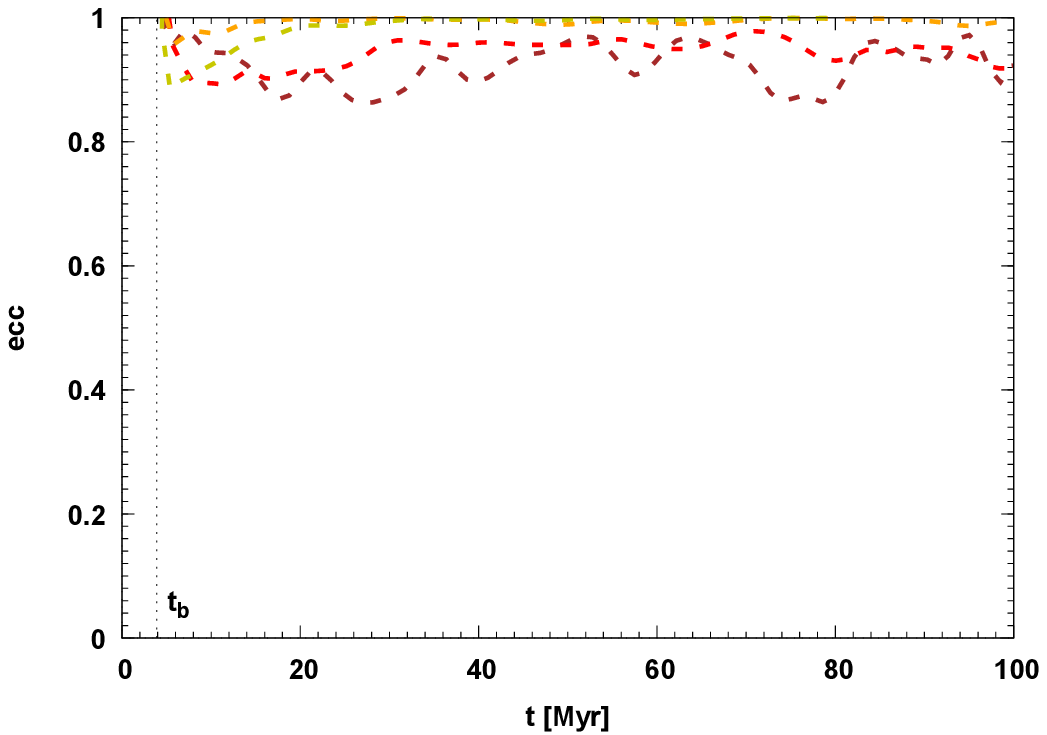}
\caption{Evolution of SMBHB separation (top), inverse semimajor axis (middle) and eccentricity (bottom) for Newtonian runs (dashed lines) with number of particles $N=100$k, randomisation seed RAND~=~1 and different mass ratios HMPs to LMPs $m_{\rm HMP}$:$m_{\rm LMP}$~=~20:1, 10:1, 5:1, 1:1 from Table~\ref{tab:num_models}. On the top panel, the solid light blue and grey lines are softening parameters for HMPs and LMPs, the solid black line is 100 Schwarzschild radii. The vertical black dashed line is binding time $t_{\rm b}$ for models (Table~\ref{tab:times}).}
\label{fig:models-hn}
\end{minipage}
\end{figure*}

\begin{table}
\caption{Timescales for models with turned on $\mathcal{PN}$ terms.}
\label{tab:times}
\centering
\begin{tabular}{ccccc}
\hline
\hline
$N$ & RAND & $t_{\rm b}$ & $t_{\mathcal{PN}\rm beg}$ & $t_{\rm merge}$ \\
    &      & Myr         & Myr                       & Myr \\
(1) & (2)  & (3)         & (4)                       & (5) \\
\hline
\hline
100k &  1  &  3.77  & 10.4 &  43.6 \\
200k &  2  &  3.90  & 10.4 &  34.7 \\
500k &  3  &  3.77  & 10.4 &  30.5 \\
500k & A.18\hyperlink{link2}{$\rm ^a$} & 5.15 & 23.5 & 40.3 \\
500k & A.25\hyperlink{link2}{$\rm ^a$} & 5.15 & 32.7 & 46.7 \\
\hline
\end{tabular}
\begin{minipage}{\linewidth}
\smallskip
NOTE:
(1) total number of particles,
(2) randomisation seed number for which we turned $\mathcal{PN}$ correction,
(3) binding binary time,
(4) time for turning $\mathcal{PN}$ correction,
(5) merging time.
\hypertarget{link2} $\rm ^a$Model A from \hyperlink{Sob21}{S21}.
\end{minipage}
\end{table}

\begin{figure*}
\centering
\includegraphics[width=0.33\linewidth]{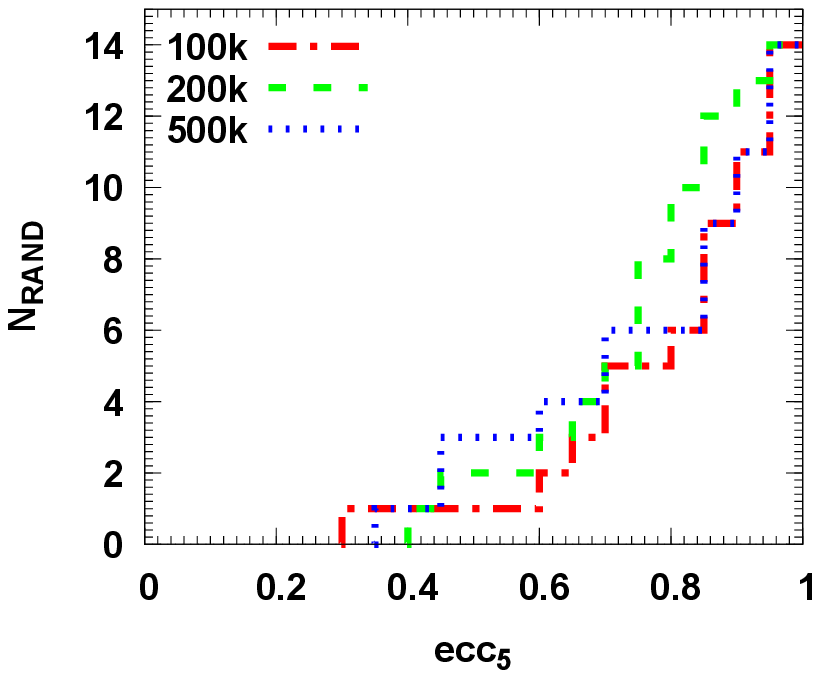}
\includegraphics[width=0.33\linewidth]{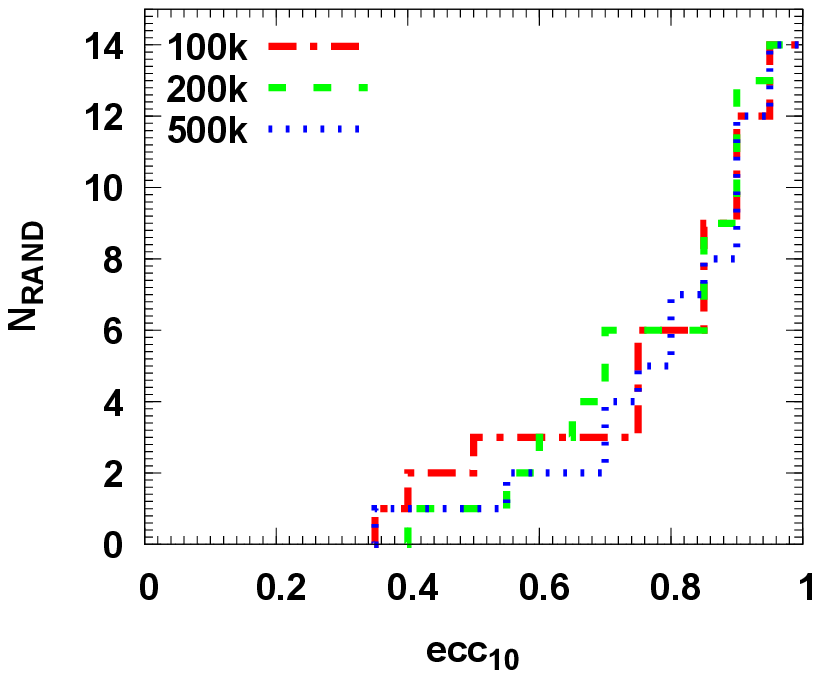}
\includegraphics[width=0.33\linewidth]{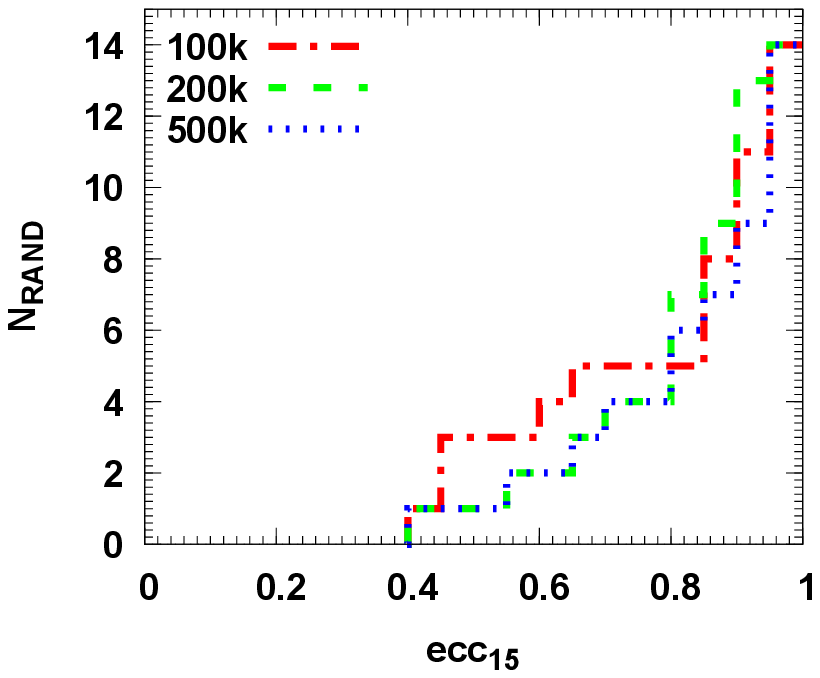}
\caption{Eccentricity cumulative distributions for numerical models with different randomisation seeds (RAND) at different times from left to right: $t$~=~5.2~Myr (5~NB; bounding time), 10.4~Myr (10~NB; turning $\mathcal{PN}$ terms) and 15.6~Myr (15~NB; forming hard binary). Colour show models with the different number of particles $N$: red -- 100k, green -- 200k, blue -- 500k.}
\label{fig:ecc-distr}
\end{figure*}

\section{Simulation results and discussion}\label{sec:results}
\subsection{Dynamical timescales}\label{sec:dyn_time}
We describe the evolution of the SMBHB by the evolution of the binary orbit's parameters, such as separation $\Delta R$, inverse semimajor axis $1/a$ and eccentricity $ecc$ (Fig.~\ref{fig:models}). As mentioned above, at time $t=0.0$~Myr the SMBH particles at initial separation $\Delta R$ are not bound. In Newtonian N-body simulations the binary forms after several passages at binding time $t_{\rm b}$ in less than 4~Myr  (Tables~\ref{tab:times}, \ref{tab:times-new}). The evolution of separation (Fig.~\ref{fig:models}, top) and inverse semimajor axis (Fig.~\ref{fig:models}, middle) show a quite good agreement for a different number of particles and initial randomisation of N-body particles' positions and velocities. This already made the results of our simulations quite independent from these purely numerical parameters. In comparison with model A (the closest model for our current research from \hyperlink{Sob21}{S21}) current basic numerical models shown an earlier ($\approx20$\%) binding time $t_{\rm b}$ (Table~\ref{tab:times}). In our set of runs, the bound binary is usually formed with a semimajor axis almost equal to the SMBHs influence radius.

For basic numerical models the eccentricity did not show any systematic dependence on the number of particles or randomisation seeds due to their very `stochastic' nature. In the basic models with 100k particles, the binaries were formed with eccentricities from $0.84$ to $0.94$. For the basic 200k runs, we get the eccentricities in the range 0.42--0.88. For the basic 500k runs we get an even wider range 0.34--0.92 (similar to in \hyperlink{Sob21}{S21}). To make our conclusion more statistically significant we performed additional Newtonian N-body simulations for $N=$~100k, 200k, 500k with different randomisation seeds and as a result totally we have 14 runs for each $N$. SMBHs orbits show a smooth trend with the orbital eccentricity higher than 0.5 (Fig.~\ref{fig:ecc-distr}). The orbital eccentricity slightly grows during the binary evolution \citep{Preto2011}. In Fig.~\ref{fig:ecc-distr} we present the cumulative eccentricity distribution for the three characteristic times (bounding time; time, when $\mathcal{PN}$ terms turn on; a time when the hard binary forming). We do not have a substantial dependence on the particle numbers $N$. Our $N$-independent wide eccentricity range (0.40--0.99) for the binaries does not really support the predictions of a more narrow eccentricity spread as an increasing number of N-body particles \citep{Rantala2017,Nasim2020}. 

After turning on the $\mathcal{PN}$ terms at time $t_{\mathcal{PN}\rm beg}=10{\rm\;T_{NB}}=10.04$~Myr all our $\mathcal{PN}$ runs show a quite short dynamical merging time $t_{\rm merge}$ comparable with obtained by \cite{Khan2016,2018ApJ...868...97K} (Table~\ref{tab:times}). Basically all three different $\mathcal{PN}$ models (100-1, 200-2, 500-3) merge in under $\approx44$~Myr (Table~\ref{tab:times}). Differences at the merging times can be explained by the strong effect of the eccentricities at the time when we turned on $\mathcal{PN}$ corrections. Previous detailed study of 20 physical and numerical models showed that merging time for central SMBHB is less than 50~Myr (for full description see \hyperlink{Sob21}{S21}). But our current binary models can merge even earlier around 31~Myr (model 500-3), which can be explained by a higher eccentricity ($\approx0.9$) at the binary formation time than in \hyperlink{Sob21}{S21}. 

\begin{figure}
\centering
\includegraphics[width=0.98\linewidth]{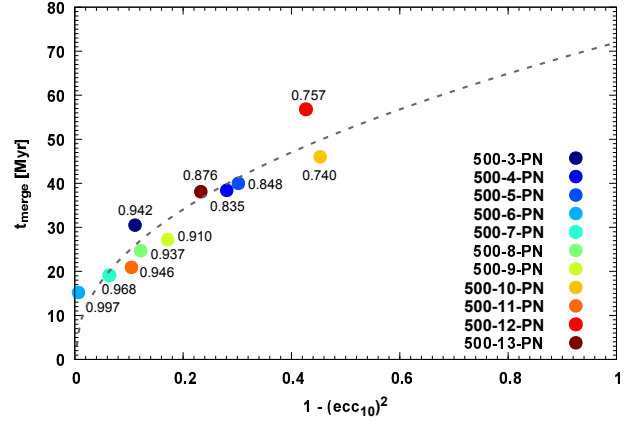}
\caption{SMBHB merging time as function of eccentricity at time $t=10.4$~Myr (10.0~NB),  when we started $\mathcal{PN}$ runs. Colour show models with different randomisation seeds RAND and numbers show the eccentricity values. Grey dashed line is fitting function (see equation~\ref{eq:fit-tmerge}).}
\label{fig:models-new}
\end{figure}

To check the merging time dependency of our PN runs from the different randomisation seeds (RAND) for the particle distributions we carry out extra 10 runs of the 500-3-PN model (Table~\ref{tab:times-new}). Before starting the extra PN runs we estimated the bounding time $t_{\rm b}\approx4$~Myr and hardening time $t_{\rm h}\approx15$~Myr for each run. The SMBHB merging time varies in a range from 15.2~Myr to 56.8~Myr and, as we expect, mainly depends on the initial eccentricity after the moment of the binary formation (Fig.~\ref{fig:models-new}). From our limited sample (totally 11 $\mathcal{PN}$ simulations) we already can conclude that the merging time can be approximated as a quite shallow function of the eccentricity:
\begin{equation}
t_{\rm merge} = {\rm A}\times\left[1-(ecc_{10})^{2}\right]^{\rm B},
\label{eq:fit-tmerge}
\end{equation}
where coefficients ${\rm A}=71.98\pm7.89$ and ${\rm B}=0.46\pm0.07$. As a basic conclusion from these extra 10 runs, we can state that even for the very small initial eccentricity the merging time has the upper limit around $\approx70$~Myr.

In the Fig.~\ref{fig:models-hn} we show the results from our extra runs with 100k particles (Table~\ref{tab:num_models}, tree bottom models), which we started to check the effect of different HMPs to LMPs mass ratios ($m_{\rm HMP}$:$m_{\rm LMP}$~=~20:1, 10:1, 5:1, 1:1). Our runs with mass prescriptions show a qualitatively similar evolution in separation, inverse semimajor axis and even eccentricity. For the inverse semimajor axis $1/a$ (Fig.~\ref{fig:models-hn}, middle) we see the trend, that is more significant at time $\approx100$~Myr. This trend strongly depends on the limit close to the 1:1 particles mass ratio and is determined by the mass of LMPs (see Table~\ref{tab:num_models} for $m_{\rm LMP}$). Because we always have a larger amount of LMPs (i.e. more interaction with the LMP particles), the binary hardening always more strongly depends on the LMPs masses. The small amount of HMPs ($\approx9\%$), in each mass prescription model, apparently is not enough for extracting sufficient energy amount during three-body encounters with the binary SMBH. For a quantitative description of this process, a detailed study of energy balance is required \citep[for example as it was made by][]{Avramov2021}.

\begin{table}
\caption{Timescales for additional numerical models with $N=500$k and different randomisation seeds.}
\label{tab:times-new}
\centering
\begin{tabular}{cccc}
\hline
\hline
RAND & $t_{\rm b}$ & $t_{\rm h }$ & \multicolumn{1}{c}{$t_{\rm merge}$} \\
     & Myr & Myr & \multicolumn{1}{c}{Myr} \\
(1) & (2) & (3) & (4) \\
\hline
\hline
4  & 4.10 & 14.04 & 38.4 \\
5  & 4.13 & 14.95 & 40.0 \\
6  & 4.10 & 14.60 & 15.2 \\
7  & 4.04 & 12.87 & 19.1 \\
8  & 4.03 & 13.40 & 24.7 \\
9  & 4.10 & 15.34 & 27.2 \\
10 & 4.16 & 14.69 & 46.0 \\
11 & 3.87 & 13.91 & 20.9 \\
12 & 4.16 & 14.04 & 56.8 \\
13 & 4.06 & 14.30 & 38.1 \\
\hline
\end{tabular}
\begin{minipage}{\linewidth}
\smallskip
NOTE:
(1) randomisation seed number for which we turned $\mathcal{PN}$ corrections,
(2) binding binary time,
(3) form hard binary time,
(4) merging time.
\end{minipage}
\end{table}

For mass prescription models the eccentricity (Fig.~\ref{fig:models-hn}, bottom) varies in a narrower range 0.85-0.99 than for basic numerical runs (Fig.~\ref{fig:models}, bottom). We do not see any strong dependence of the binary initial eccentricity from the LMPs particles individual masses. Lines for different models are very often overlapping (crossing). We can just note that models with higher mass ratios (20:1, 10:1) have some kind of `bumps'. This can indicate the interaction with the particular HMP.  Even if their number are much lower compare to the LMP such a small amount of high mass field particles can play a significant role in the binary eccentricity behaviour.

\subsection{Gravitational waves}\label{sec:gw}
For our model with maximum $N$ and turning on $\mathcal{PN}$ terms ($N=500$k, RAND~=~3, $m_{\rm HMP}$:$m_{\rm LMP}$~=~10:1) we also calculated the expected amplitude-frequency picture for SMBHB merging in NGC~6240. For the simple waveform calculation we used the GW quadrupole term expressions from \cite{Kidder1995} \citep[also see][]{Brem2013,Sobolenko2017}: 
\begin{equation}
h^{ij} = \frac{2G\mu}{D_{\rm L}c^4}~\left[Q^{ij}+P^{0.5}Q^{ij}+PQ^{ij}+P^{1.5}Q^{ij}+~...\right],
\end{equation}
where $P$ is a correction term for corresponding $\mathcal{PN}$ order, $\mu$ is the reduced mass, $D_{\rm L}$ is the luminosity distance between the origin of the reference frame and the source, and $Q^{ij}$ is the quadrupole term. The last one can be written in the form:
\begin{equation}
Q^{ij}=2\left[\varv^{i}\varv^{j}-\frac{GM_{\rm BH12}}{r}n^{i}n^{j}\right],
\end{equation}
where $\varv^{i}$ and $n^{i}$ are the relative velocity and normalised position vectors in this reference frame respectively.

For illustrative purpose we did not highly accurate model waveforms and neglected the higher order terms. In this assumption we calculated the tensor in the source frame simply by:
\begin{equation}
h^{ij}\approx\frac{4G\mu}{D_{\rm L}c^{4}}\left[\varv^{i}\varv^{j}-\frac{GM_{\rm BH12}}{r}n^{i}n^{j}\right].
\end{equation}
For the sake of simplicity, we choose the virtual detector to be oriented such that the coordinate axes coincide with the source frame. It allowed us did not make any coordinate transformations. We computed $h_{+}$ and $h_{\times}$ from $h^{ij}$, which gave the relevant measurable strains in `+' and `$\times$' polarisations \citep{Brem2013,Sobolenko2017}.

\begin{figure*}
\centering
\includegraphics[width=0.47\linewidth]{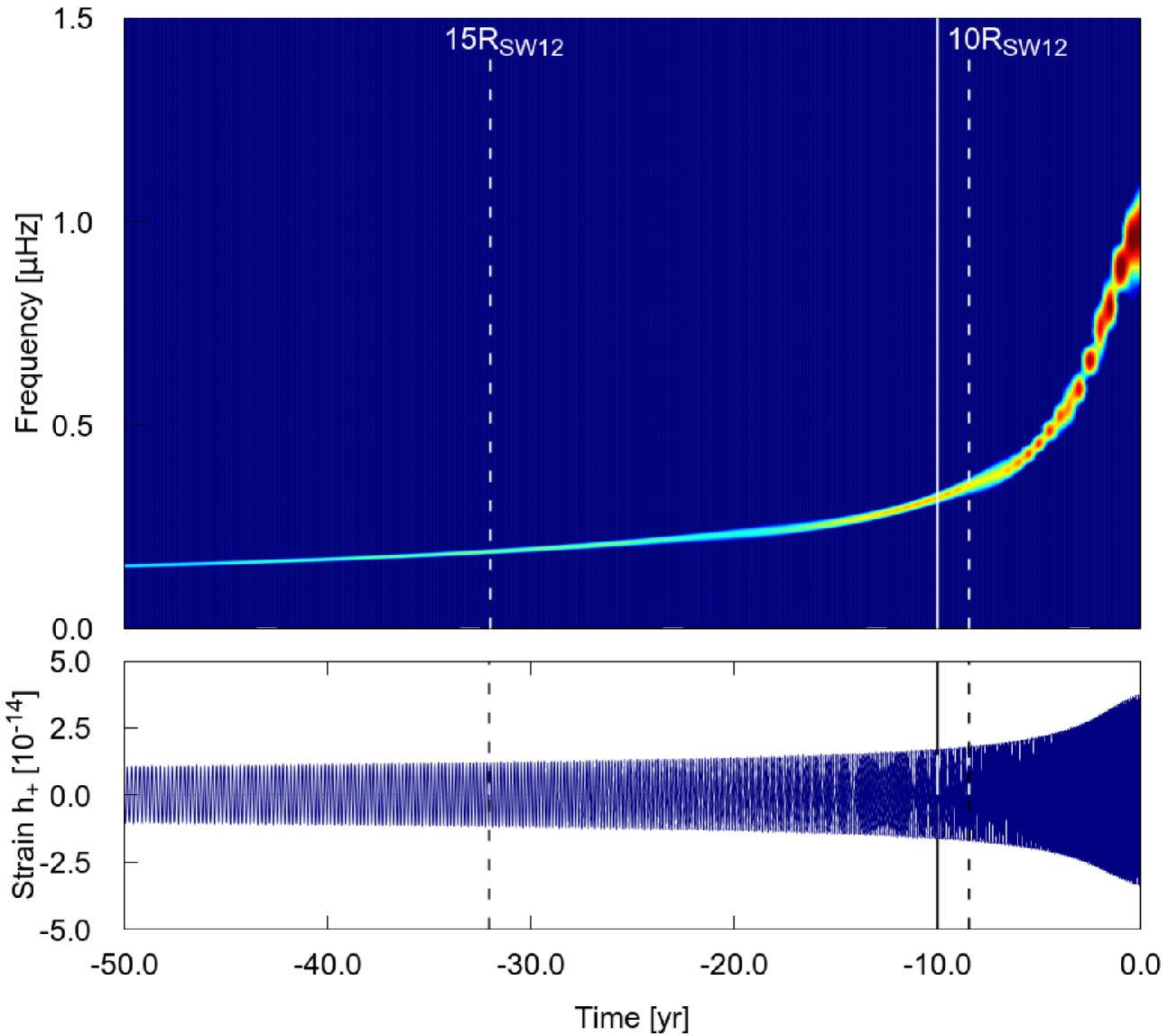}
\includegraphics[width=0.52\linewidth]{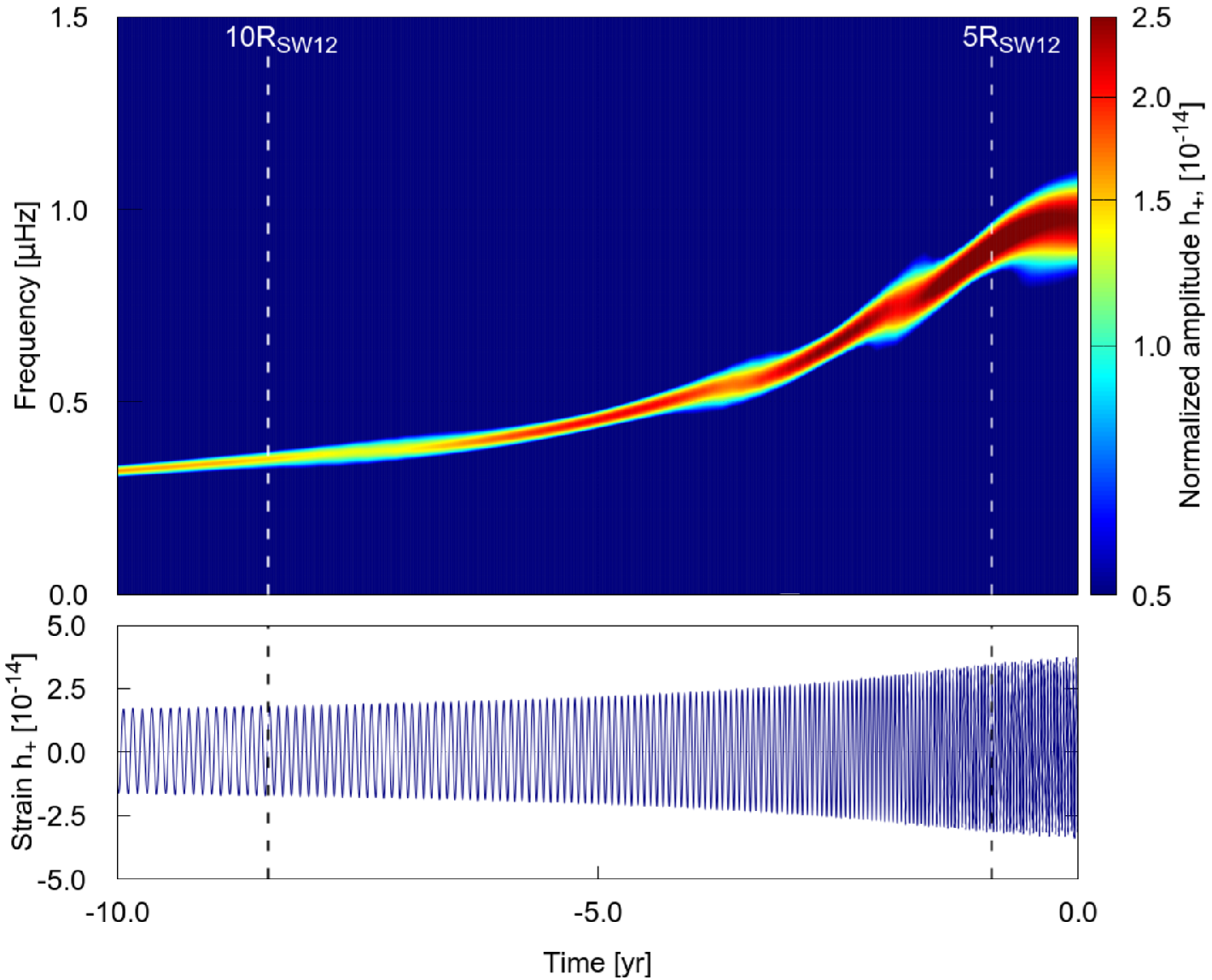}
\caption{Time-frequency representations (top) of the strain data (bottom) for predicted gravitational waveforms of $h_{+}$ polarisation from SMBHB merging at NGC~6240 ($D_{\rm L}=111.2$~Mpc) for the last 50~yr (left) and last 10~yr (right). Major merging is represented by binary component with masses $1.36\times10^{9}\rm\;M_{\odot}$ and $6.8\times10^{8}\rm\;M_{\odot}$ and corresponding mass ratio 2:1. The final separation (due to our $\mathcal{PN}$ routine) is 0.75~mpc. The solid vertical line on the left panel indicates the last 10~yr of merging. Dashed vertical lines from left to right indicate binary separation 15, 10 and 5 Schwarzschild radii respectively.}
\label{fig:GW}
\end{figure*}

The standard resolution for our $\mathcal{PN}$ runs was $1.3\times10^{5}$~years. We extracted the SMBH particles data (positions \& velocities) from the last available $\mathcal{PN}$ model's snapshot to calculate the final stage of the SMBH merger (up to $\approx 4 R_{\rm SW12}$) with the high resolution.  Using these particle data, we followed only the two SMBHs dynamical $\mathcal{PN}$ evolution. For this purpose, we used our highly accurate two-body Hermite integrator. We run these separate simulations with the maximum possible accuracy, keeping at minimum 100 points per SMBH particles orbital integration, which give us time resolution up to $\sim3$ days.

The calculated waveforms for $h_{+}$ polarisation and amplitude-frequency picture from the final phase of our model runs (last 50~years and zoomed last 10~years evolution before the merger) we present in Fig.~\ref{fig:GW}. It is worth noting that $\mathcal{PN}$ approximation works well for describing the early inspiral SMBHBs, and numerical relativity and perturbation theory should be used for full waveforms picture of merging event and ringdown \citep[for reference see][]{LeTiec2014}. Obtained frequencies for merging events from such high mass SMBHs ($\sim10^{8-9}\rm\; M_{\odot}$) at such distances ($D_{\rm L}=111.2$~Mpc) lay on sensitive curve of current and future pulsar timing array (PTA) consortium's: European PTA \citep[EPTA,][]{Kramer2013}, Parkes PTA \citep[PPTA,][]{Hobbs2013}, North American Nanohertz Observatory for Gravitational Waves \citep[NANOGrav,][]{Ransom2019}, which collectively form International PTA \citep[IPTA,][]{Manchester2013}. Such detection of individual SMBHBs merging and GWs stochastic background \citep[see the recent NANOGrav 12.5~yr data set results at][]{Arzoumanian2020} will be strong evidence of the possibility of SMBHs binding, their reaching sub-pc scale, merging and emitting GWs.

\section{Conclusions} \label{sec:conc}

In this paper, we investigated the X-ray properties of dual AGN in NGC~6240 using \textit{Chandra} observations in the 0.5--7.5~keV and performed numerical N-body simulations based on the results of the corresponding spectral analysis. The main conclusions of this study can be summarised as follows.
\begin{enumerate}
    \item We performed X-ray analysis of the combined spectrum from four \textit{Chandra} observations of NGC~6240 with resulting exposure of $~480$~ks for each of two active nuclei. These spectra demonstrated individual Fe K$\alpha$ emission lines with observational energies $E_{\rm S}=6.39_{-0.02}^{+0.01}$~keV and $E_{\rm N}=6.41_{-0.02}^{+0.01}$~keV  with corresponding line widths $\sigma_{\rm S}=0.05_{-0.03}^{+0.04}$~keV and $\sigma_{\rm N}=0.05^{+0.01}_{-0.02}$~keV for South and North nuclei respectively. 
    \item We estimated the dynamical mass for these nuclei as $M_{\rm dyn} \approx 2.04\times10^{11}\rm\;M_{\odot}$ from X-ray analysis assuming that obtained energy shift caused by the relative motion of the two nuclei at the late stage. Accepting that this mass represents the mass of bulge, we estimated SMBHB mass as $M_{\rm BH12} \approx 2.04\times10^{9}\rm\;M_{\odot}$. This value is comparable with estimations by other authors \citep{Medling2011,Kollatschny2020}.
    \item Based on the estimated bulge mass and maximum projected separation $\Delta  R=1$~kpc of the central SMBHB we constructed a physical model of the merging system. Using this physical model we made twelve basic numerical models' realisations with different particles number $N=100$k, 200k, 500k. To obtain the merging time we run Newtonian and  $\mathcal{PN}$ N-body models (up to $2.5\mathcal{PN}$ term). As a basic code, we used our own direct N-body \hyperlink{link1}{\PGPU} code with 4$^{\rm th}$ order Hermite integration scheme and individual timesteps for particles. 
    \item All basic Newtonian simulations showed a very good alignment in inverse semimajor axis evolution. From these runs, we concluded the independence of our SMBH binary hardening results on the initial number of particles (100k, 200k, and 500k) and randomisation for particles' positions and velocities. The eccentricity did not show any systematic dependence neither on the number of particles nor randomisation seeds due to its very `stochastic' nature. 
    \item To make our conclusions more statistically significant we performed extra Newtonian N-body simulations for $N=$~100k, 200k, 500k with different randomisation seeds. For extra simulations, eccentricity also did not show any substantial dependence on the particle numbers $N$. Our $N$-independent wide eccentricity range (0.40--0.99) for the binaries does not support the predictions \citep{Rantala2017,Nasim2020} of a more narrow eccentricity spread as an increasing number of N-body particles.
    \item To estimate the merging time for a central SMBHB we combined the basic Newtonian and $\mathcal{PN}$ numerical models. The obtained merging times lay in a range from 15~Myr to 57~Myr, which is in quite good agreement with our previous results \citep{Sobolenko2016,Sobolenko2021}. The extra ten $\mathcal{PN}$ Newtonian and $\mathcal{PN}$ models with $N=500$k and different randomisation seeds for the particle distributions show also a quite similar result. Based on the numerical approximation of the merging time as a function of SMBH binary eccentricity we can conclude that even for the possibly very small initial eccentricity the merging time anyway has an upper limit around $\approx70$~Myr.
    \item Implementing relativistic $\mathcal{PN}$ approximation up to $2.5\mathcal{PN}$ terms allowed us to follow the SMBHB evolution till the mpc scale. We obtained the waveforms and amplitude-frequency maps for the last 50 and 10 years for the SMBHB system in interacting galaxy NGC~6240. Such SMBHBs merging events can be observed in the current and future PTA campaigns.
\end{enumerate}
The presented complete research from observation analysis to numerical modelling gives us a powerful key for detailed investigation of the complex objects such as double/multiple AGN systems at different merging stages.

\section*{Acknowledgements}
The authors thank for the anonymous referee for the fruitful comments and corrections of our mistakes/typos in the manuscript. 
We believe, that her/his useful  comments greatly improved the final version of our paper. 

The authors gratefully acknowledge the Gauss Centre for Supercomputing (GSC) e.V. (\url{www.gauss-centre.eu}) for funding this project by providing computing time through the John von Neumann Institute for Computing (NIC) on the GCS Supercomputers JURECA and 
JUWELS at J{\"u}lich Supercomputing Centre (JSC).
This research has made use of data obtained from the Chandra Data Archive and the Chandra Source Catalog, and software provided by the Chandra X-ray Center (CXC) in the application packages CIAO and Sherpa.

OK is grateful to Dr. O. Torbaniuk for useful discussions and comments and Dr. I. Vavilova for helpful remarks.
OK  acknowledge support by the budgetary program `Support for the development of priority fields of scientific research' (CPCEL 6541230), project No.~10-F, and the Target Complex Program of Scientific Space Research of the NAS of Ukraine.
OK thanks the Astronomical Observatory of the Jagiellonian University for support during the International Summer Student Internship, where a part of this work was done.

The work of VM was supported by the Polish NSC grant 2016/22/E/ST9/00061.

MS acknowledges the support under the Fellowship of the National Academy of Sciences of Ukraine for young scientists 2020-2022. 
PB and MS acknowledge support by the Volkswagen Foundation in Germany under the grant No.~97778. The work of PB was also supported by the Volkswagen Foundation under the special stipend No.~9B870 (2022).

PB and BS also acknowledges the support from the Science Committee of the Ministry of Education and Science of the Republic of Kazakhstan (Grant No. AP08856149).
PB thanks the support by Ministry of Education and Science of Ukraine under the France - Ukraine collaborative grant M2-16.05.2022. 
PB express acknowledge the support by the National Academy of Sciences of Ukraine under the Main Astronomical Observatory GPU computing cluster project No.~13.2021.MM.

The work of PB, AV and MS was supported under the special program of the NRF of Ukraine `Leading and Young Scientists Research Support' -- `Astrophysical Relativistic Galactic Objects (ARGO): life cycle of active nucleus', No.~2020.02/0346.

BS acknowledges the Nazarbayev University Faculty Development Competitive Research Grant Program No~11022021FD2912~(ssh2022007).

\section*{Data Availability}
The data underlying this article will be shared on reasonable request to the corresponding author.

\bibliographystyle{mnras}
\bibliography{NGC6240-Chandra} 

\bsp	
\label{lastpage}
\end{document}